\documentclass[11pt,a4paper]{article}
\pdfoutput=1
 \usepackage{jheppub}
\allowdisplaybreaks
\usepackage{graphicx,psfrag}
\usepackage{bm}
\usepackage{mathbbol,verbatim}
\usepackage{booktabs,siunitx}
\usepackage{booktabs}
\usepackage{multirow}
\usepackage{siunitx}
\usepackage{slashed}
\usepackage{graphics}
\usepackage{color,ulem}
\usepackage{multirow}
\usepackage{tikz}
\usetikzlibrary{trees}
\usetikzlibrary{decorations.pathmorphing}
\usetikzlibrary{decorations.markings}
\usetikzlibrary{decorations, decorations.markings, decorations.pathmorphing, arrows, graphs, shapes.geometric, snakes}
\usetikzlibrary{arrows}

\def\beq{\begin{equation}}
\def\eeq{\end{equation}}
\newcommand{\beqa}{\begin{eqnarray}} 
\newcommand{\eeqa}{\end{eqnarray}}
\newcommand{\barr}{\begin{array}}
\newcommand{\earr}{\end{array}}

\def\gs{\mathrel{
   \rlap{\raise 0.511ex \hbox{$>$}}{\lower 0.511ex \hbox{$\sim$}}}}
\def\ls{\mathrel{
   \rlap{\raise 0.511ex \hbox{$<$}}{\lower 0.511ex \hbox{$\sim$}}}}
   \def\beq{\begin{equation}}
\def\eeq{\end{equation}}
\def\bea{\begin{equation}}
\def\eea{\end{equation}}

\def\la{\label}

\usepackage{cancel}

\def\lapp{\mathrel{\rlap{\raise.5ex\hbox{$<$}}
                    {\lower.5ex\hbox{$\sim$}}}}
\def\gapp{\mathrel{\rlap{\raise.5ex\hbox{$>$}}
                    {\lower.5ex\hbox{$\sim$}}}}

\usepackage{graphicx,psfrag}
\usepackage{bm}
\usepackage{mathbbol,verbatim}
\usepackage{booktabs,siunitx}
\usepackage{booktabs}
\usepackage{multirow}
\usepackage{siunitx}
\usepackage{slashed}
\usepackage{graphics}
\usepackage{color,ulem}
\usepackage{multirow}
\usepackage{tikz}
\usetikzlibrary{trees}
\usetikzlibrary{decorations.pathmorphing}
\usetikzlibrary{decorations.markings}
\usetikzlibrary{decorations, decorations.markings, decorations.pathmorphing, arrows, graphs, shapes.geometric, snakes}
\usetikzlibrary{arrows}

\newcommand{\wbp}{W^+}
\newcommand{\wbm}{W^-}
\newcommand{\nub}{\overline{\nu}}
\newcommand{\lp}{\ell^+}
\newcommand{\lm}{\ell^-}

   \def\beq{\begin{equation}}
\def\eeq{\end{equation}}
\def\bea{\begin{equation}}
\def\eea{\end{equation}}

\def\la{\label}

\def\be{\begin{equation}}
\def\ee{\end{equation}}
\def\bea{\begin{eqnarray}}
\def\eea{\end{eqnarray}}
\usepackage{cancel}

\def\lapp{\mathrel{\rlap{\raise.5ex\hbox{$<$}}
                    {\lower.5ex\hbox{$\sim$}}}}
\def\gapp{\mathrel{\rlap{\raise.5ex\hbox{$>$}}
                    {\lower.5ex\hbox{$\sim$}}}}

\begin{document}

\preprint{
\begin{flushright} 
OSU-HEP-17-02\\
IP/BBSR/2017-8
\end{flushright} 
}

\title{Neutrino Mass Generation at TeV Scale and New Physics Signatures from Charged Higgs at the LHC for Photon Initiated Processes}
\author[a,b]{Kirtiman Ghosh,}
\author[a,c,d]{Sudip Jana,}
\author[d]{and S. Nandi}

\affiliation[a]{Institute of Physics, Sachivalaya Marg, Sainik School Post, Bhubaneswar 751005, India}
\affiliation[b]{Homi Bhabha National Institute, Training School Complex, Anushakti Nagar,
Mumbai 400085, India}
\affiliation[c]{Theoretical Physics Department, Fermilab, Batavia, IL 60510, USA}
\affiliation[d]{ Department of Physics and Oklahoma Center for High Energy Physics,
Oklahoma State University, Stillwater, OK 74078-3072, USA. }

\emailAdd{kirti.gh@gmail.com}
\emailAdd{sudip.jana@okstate.edu}
\emailAdd{s.nandi@okstate.edu}

\date{\today}

\abstract{We consider the collider phenomenology of  a simple extension of the Standard Model (SM), which consists of an EW isospin $3/2$ scalar, $\Delta$ and a pair of EW isospin $1$ vector like fermions, $\Sigma$ and $\bar{\Sigma}$, responsible for generating tiny neutrino mass via the effective dimension seven operator. This scalar quadruplet with hypercharge Y = 3 has a plethora of implications at the collider experiments. Its signatures at TeV scale colliders are expected to be seen, if the quadruplet masses are not too far above the electroweak symmetry breaking scale. In this article, we study the phenomenology of multi-charged  quadruplet scalars. In particular, we study the multi-lepton signatures at the Large Hadron Collider (LHC) experiment, arising from the production and decays of triply and doubly charged scalars. We studied Drell-Yan (DY) pair production as well as pair production of the charged scalars via photon-photon fusion. For doubly and triply charged scalars, photon fusion contributes significantly for large scalar masses. We also studied LHC constraints on the masses of doubly charged scalars in this model. We derive a lower mass limit of 725 GeV on doubly charged quadruplet scalar.}

\maketitle
\section{Introduction}
Evidence of physics beyond the Standard Model (SM) have essentially come from one of the most important discoveries  {namely,} the   {discovery} of non-zero tiny neutrino masses. In this paper, we consider  {a model}  {which} naturally  {accommodate} small neutrino masses arising from  dimension-7 operators.  In order to realize TeV scale seesaw mechanism, the model includes  a scalar quadruplet and a pair of vector-like fermion triplets.   {The characteristic signatures of this model at the hadron collider experiments like the Large Hadron Collider (LHC),  arise from the production and decay of the triply- and doubly- charged scalars of the scalar quadruplet.}   {In particular,} the observation of a triply-charged scalar  {at the LHC}  would establish this type of  seesaw mechanism as the most promising framework for generating neutrino masses.  {The charged scalars, in the framework of this model, dominantly decay into charged SM leptons and thus, result into tantalizing same-sign multi-lepton final states at the LHC.}

 {The ATLAS and CMS collaborations of the LHC  experiment have already performed dedicated searches \cite{ATLAS:2016pbt,ATLAS:2014kca,ATLAS:2012hi,CMS:2016cpz,Chatrchyan:2012ya} for like-sign dileptons as a signature of a doubly charged scalar ($\Delta^{\pm\pm}$). In absence of any significant deviation of data from the SM prediction, bounds are imposed on the mass of $\Delta^{\pm\pm}$ as a function of its decay into lepton pairs. For example, a search \cite{ATLAS:2014kca} for anomalous production of like-sign lepton (electron and muon only) pairs, arise from the production and decay of a doubly charged scalar, $\Delta^{\pm\pm}$, was performed by ATLAS collaboration with 20.3 fb$^{-1}$ of 8 TeV proton-proton collision data. Assuming 100\% branching ratio (BR) of $\Delta^{\pm\pm}$ into a pair of leptons of a given flavor, a 95\% CL lower limit of 465--550 (370--435) GeV (depending on the lepton flavour) in the context of left-right symmetry was obtained on the mass of left-(right-)handed $\Delta^{\pm\pm}$. CMS collaboration  \cite{CMS:2016cpz,Chatrchyan:2012ya} with 4.93 fb$^{-1}$(19.7 fb$^{-1}$) integrated luminosity of collected data at the LHC with 7(8) TeV center of mass energy had excluded doubly charged scalar mass below 169--395 (251--530) GeV. The ranges correspond to 100\% BR into different combinations of same-sign dilepton flavours in the final state, {\em i.e.,} $e^{\pm}e^{\pm}, e^{\pm}\mu^{\pm}, e^{\pm}\tau^{\pm}, \mu^{\pm}\mu^{\pm}, \mu^{\pm}\tau^{\pm}$ and $\tau^{\pm}\tau^{\pm}$. More stringent limits \cite{ATLAS:2016pbt} {\em i.e.,} 380 (530) GeV for $\Delta^{\pm \pm}_{R(L)}$ decaying into a pair of electrons with 50\% BR, are now available from the LHC with 13 TeV center of mass energy and 13.9 fb$^{-1}$ integrated luminosity.}

 {Quadruplet scalars, being charged under the SM gauge group, couple to photon and the SM electroweak (EW) gauge bosons ($Z$ and $W^{\pm}$). Therefore, these scalars are produced in pairs at the LHC from quark anti-quark initial state via a $\gamma/Z/W^\pm$ exchange in the $s$-channel namely, via the Drell-Yan (DY) process. The experimental limits, discussed in the previous paragraph, are obtained assuming DY pair production of doubly charged scalars. However, charged scalars are also produced via $t(u)$-channel photon-photon fusion process. Photon density\footnote{The inclusion of the photon as a parton inside the proton, with an associated parton distribution function (PDF) is required to include next-to-leading order (NLO) QED corrections. Since $\alpha_S^2$ is of the same order of magnitude as $\alpha_{EM}$ and in the era of precision phenomenology at the LHC when the PDFs are already determined upto NNLO in QCD, consistency of calculations require PDFs which are corrected atleast upto NLO QED.} being significantly smaller than the quark and gluon densities, photon fusion contribution to the pair-production of charged scalars was neglected in the literature \cite{earlylhc, earlylhc2} as well as by the experimental groups \cite{ATLAS:2016pbt,ATLAS:2014kca,ATLAS:2012hi,CMS:2016cpz,Chatrchyan:2012ya}. However, photon coupling to a pair of charged scalar being proportional to the charge of the scalar, parton level photon fusion cross-sections are enhanced by a factor of $2^4$ and $3^4$ for the doubly and triply charged scalars, respectively. Moreover, photon fusion being a $t(u)$-channel process, falls slowly with parton center of mass energy ($\sqrt{\hat s}$) compared to the $s$-channel DY process. Therefore, for larger masses of doubly and triply charged scalars, photon fusion production could be significant compared to the conventional DY production. }

 {In this work, we have performed a comparative study of DY and photon fusion pair-production of multi charged scalars at the LHC with 13 TeV center of mass energy. It was shown for the first time,  that  for large scalar masses {the}  photo productions of triply and doubly charged   {scalars} via the  photon fusion  contribute at a level comparable to the DY-productions. As a consequence, all the LHC search results for charged   {scalars} change dramatically after photon initiated processes.  {In the context of present model, we obtained bound on the mass of doubly charged quadruplet scalar from the LHC doubly charged scalar search results and hence, excluded some parts of parameter space. We also studied the production and decay of triply charged scalars at 13 TeV LHC.}

This paper is organized as follows. In section \ref{sec:2}, we discuss about the model and neutrino masses. In section \ref{sec:3}, we  {briefly} discuss the production and decay modes of   {doubly and triply charged scalars}, derive the exclusion limit on the doubly charged scalar mass and hence, on the parameter space, from the LHC 13 TeV results. We also analyze the characteristic collider signatures of these scalars at the future runs of the LHC. In the last part of section 3, we briefly discussed the possible collider signatures of the triplet leptons ($\Sigma$ and $\bar \Sigma$) which are an integral part of this model. We finally conclude in section \ref{sec:4}.

 

\section{Model and Formalism} \label{sec:2}
 {In order to realize see-saw mechanism for generating tiny neutrino masses}, in addition to the  {usual} SM  {matter fields,}  {the model \cite{Babu:2009aq}}  {includes} two vector-like $SU(2)_L$ triplet leptons ($\Sigma $ and $\bar{\Sigma}$)  {and}  an isospin $3/2$ scalar ($\Delta$)  {in the framework of the SM gauge symmetry : $SU(3)_{C}\times SU(2)_{L}\times U(1)_{Y}$}. The particle contents along with their quantum numbers are shown in the Table~\ref{Table1}.  \\
\begin{table}[htb]
\begin{center}
\begin{tabular}{c c c}
\hline 
\hline
& \textbf{$SU(3)_{C} \times SU(2)_{L} \times U(1)_{Y}$}\\  \hline \hline
\small  \textbf{Fermions} : &${\begin{pmatrix} u \\ d \end{pmatrix}}_L\sim(3,2,\frac{1}{3}), u_R\sim (3,1,\frac{4}{3}), d_R \sim (3,1,-\frac{2}{3})$ \\ \\
&$ {\begin{pmatrix} \nu_e \\ e \end{pmatrix}}_L\sim (1,2,-1), e_R\sim (1,1,-2)$ \\ \\ &$ {\Sigma \equiv \begin{pmatrix} \Sigma^{++} \\ \Sigma^{+} \\ \Sigma^{0} \end{pmatrix}}\sim (1,3,2)$, $ {\bar{\Sigma} \equiv \begin{pmatrix} \bar{\Sigma}^{0} \\ \bar{\Sigma}^{-} \\ \bar{\Sigma}^{--} \end{pmatrix}}\sim (1,3,-2)$ \\ \\
\small \textbf{Gauge} :  & $G^\mu_{a,a=1-8}, A^\mu_{i, i=1-3}, B^\mu$ \\ \\ \\
\small \textbf{Higgs} : & $H\equiv{\begin{pmatrix} \phi^{+} \\ \phi^{0} \end{pmatrix}}\sim(1,2,1)$, $\Delta\equiv{\begin{pmatrix} \Delta^{+++} \\ \Delta^{++} \\ \Delta^{+} \\\Delta^{0}\end{pmatrix}}\sim(1,4,3)$.\\ 
\hline
\hline
\end{tabular}
\\
\end{center}
\caption{Fermion, gauge and Higgs contents of the model.}
\label{Table1}
\end{table}


The most general renormalizable  scalar potential  consistent with   {scalar spectrum of this} model is given by,
\begin{equation}
\begin{split}
V ( H, \Delta)= \mu_H^{2}H^{\dagger}H + \mu_\Delta^{2}\Delta^{\dagger}\Delta +   \frac{\lambda_{1}}{2}(H^{\dagger}H)^{2} + \frac{\lambda_{2}}{2}(\Delta^{\dagger}\Delta)^{2} \\+  \lambda_{3}(H^{\dagger}H)(\Delta^{\dagger}\Delta) + \lambda_{4}(H^{\dagger}\tau_{a}H)(\Delta^{\dagger}T_{a}\Delta)  + \lbrace\lambda_{5}H^{3}\Delta^{\star} + h.c. \rbrace ,
\end{split}
\end{equation}
where  $\tau_{a}$ and $T_{a}$   {are} the generators of $SU(2)$ in the doublet  {and} four-plet representations, respectively. \\

\begin{figure}[htb]
\begin{center}
\includegraphics[height=8cm,width=10cm]{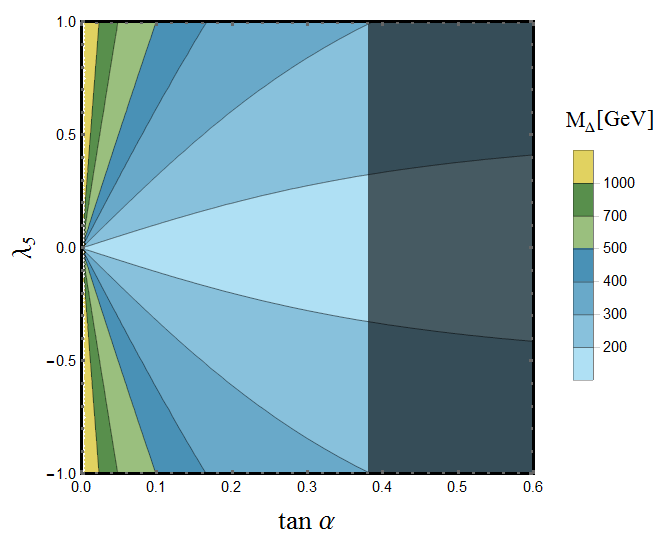}
 \caption{Contour plot for $M_{\Delta}$ in $\lambda_{5}-\tan{\alpha}$ plane. Mass scale for different color shaded regions is shown in the right side of the figure. Black shaded zone is excluded by current experimental limit.} 
 \label{p1}
 \end{center}
 \end{figure}

The EW symmetry is broken spontaneously once the neutral component of the scalar doublet (the SM Higgs doublet, $H$)  acquires the vacuum expectation value (VEV), $v_{H}$. As was shown in \cite{Babu:2009aq}, even with positive ${\mu_{\Delta}}^2$, due to the $\lambda_5$ term in the potential,  the neutral component of $\Delta$ acquires an induced VEV  at the tree level, $v_{\Delta} = - \lambda_5 v_H^3 / M_{\Delta}^2$. Constraint on $v_{\Delta}$ (in particular, $v_\Delta<2$ GeV) arises from the experimental limit \cite{Beringer:1900zz} on the $\rho$ parameter  which gets modified as $\rho \approx (1-6 v_{\Delta}^{2}/v_{H}^{2})$.  The masses of neutral ($M_\Delta$) and charged ($M_{\Delta^i}$) component of isospin-3/2 scalars are given by \cite{Babu:2009aq,gjn}
\begin{eqnarray}
 M_{\Delta}^{2} &=& \mu_\Delta^{2} + \lambda_{3}v_{H}^{2} + \frac{3}{4}\lambda_{4}v_{H}^{2},\nonumber\\
M_{\Delta^i}^{2} &=& M_{\Delta}^{2} - q_{i}\frac{\lambda_{4}}{2}v_{H}^{2},
\label{spectrum}
\end{eqnarray}
where $q_{i}$ is the (non-negative) electric charge of the respective field. The mass splittings are equally spaced and there are two possible mass orderings. For $\lambda_{4}$ positive,
we have the ordering $M_{\Delta^{+++}} < M_{\Delta^{++}} < M_{\Delta^{+}} < M_{\Delta^{0}}$ and for $\lambda_{4}$ negative, we have the ordering $M_{\Delta^{+++}} > M_{\Delta^{++}} > M_{\Delta^{+}} > M_{\Delta^{0}}$.
Due to the $\lambda_5$ term in the potential, there will be small mixing ($\alpha$) between SM Higgs and $\Delta$ and it is given by
\begin{equation}
\tan{2\alpha}=\dfrac{3\lambda_{5}v^{2}_{H}}{\sqrt{\left(M_{\Delta} ^{2}-M_{h}^{2} \right)^{2}-9\lambda_{5}^{2}v_{H}^{4}}}.
\end{equation} 
A contour plot for the mass $M_{\Delta}$ in mixing-coupling plane is shown in Figure \ref{p1}. The mixing parameter $\alpha$ can be constrained from current experimental limit \cite{ATLAS:2016nke} and it is shown by black shaded zone in Figure \ref{p1}. 

\subsection{Origin of Neutrino Masses}

Neutrino masses arise \cite{Babu:2009aq}  from the  {following Yukawa interactions involving the heavy leptons $\Sigma$ and $\bar\Sigma$:}
\beq 
{\cal L_{\nu-{\rm mass}}} = Y_i\overline{L^C_{ia}}\epsilon^{aa^\prime}\Sigma_{a^\prime b}H^{*b} +
\overline{Y}_i\overline{L^C_{ia}}\epsilon^{aa^\prime}\Delta_{a^\prime b c} \epsilon^{bb^\prime}\epsilon^{cc^\prime}\bar\Sigma_{b^\prime c^\prime} +M_{\Sigma }\overline{\Sigma^C_{ab}}\epsilon^{aa^\prime}\epsilon^{bb^\prime}\bar\Sigma_{a^\prime b^\prime}  + h.c., \la{Lsig} 
\eeq
where $Y_i,~\overline{Y}_i$ are Yukawa couplings, $i$ is the generation index and $C$ denotes charge conjugation. In the above, we have used the tensor notation (see e.g. Ref.~\cite{georgi}) to write down the triplets and quadruplet. $a,~b$ and $c$ are $SU(2)$ indices which are summed over from 1 to 2. $\epsilon^{ab}$ is a anti-symmetric rank-2 tensor with $\epsilon^{12}=1$ whereas, $\Sigma_{ab}(\bar\Sigma_{ab})$ and $\Delta_{abc}$ are total symmetric rank-2 and rank-3 tensors, respectively with the following definitions:
\begin{itemize}
\item $\Sigma_{11}=\Sigma^{++}$, $\Sigma_{12}=\Sigma^{+}/\sqrt 2$ and $\Sigma_{22}=\Sigma^{0}$ 
\item $\bar\Sigma_{11}=\bar\Sigma^{0}$, $\bar\Sigma_{12}=\bar\Sigma^{-}/\sqrt 2$ and $\bar\Sigma_{22}=\bar\Sigma^{--}$ 
\item $\Delta_{111}=\Delta^{+++}$, $\Delta_{112}=\Delta^{++}/\sqrt{3}$, $\Delta_{122}=\Delta^{+}/\sqrt{3}$ and $\Delta_{222}=\Delta^{0}$
\end{itemize}
\noindent The fermion bi-linear term in Eq.~\ref{Lsig} involving $SU(2)$ triplet fermions ($\Sigma$ and $\bar\Sigma$) is required to generate lepton number violation and hence, Majorana masses for the neutrinos. The violation of lepton number is directly reflected from the mass insertion in the propagator of the tree level as well as 1-loop diagrams in Fig.~\ref{fig:loop}. Integrating out the $\Sigma , \overline{\Sigma
}$ fermions, one obtains an effective  {dimension-5} neutrino mass operator \cite{Babu:2009aq,earlylhc}
\bea \la{neumass}
{\cal L}_{\rm eff} = -{(Y_i \overline{Y}_j +  Y_j \overline{Y}_i) L_iL_j H^* \Delta \over M_\Sigma}+h.c.
\eea
The tree level diagram generating this operator is shown in Figure~\ref{fig:loop}(top panel).  {On the other hand, the 1-loop diagrams in the bottom panel of figure \ref{fig:loop} result into dimension-5 operator which also contributes to the neutrino masses.} The detailed structure of the Yukawa interactions are given in \cite{Babu:2009aq,earlylhc}.
\begin{figure}[htb]
\begin{center}
$$
\includegraphics[height=6.3cm,width=6cm]{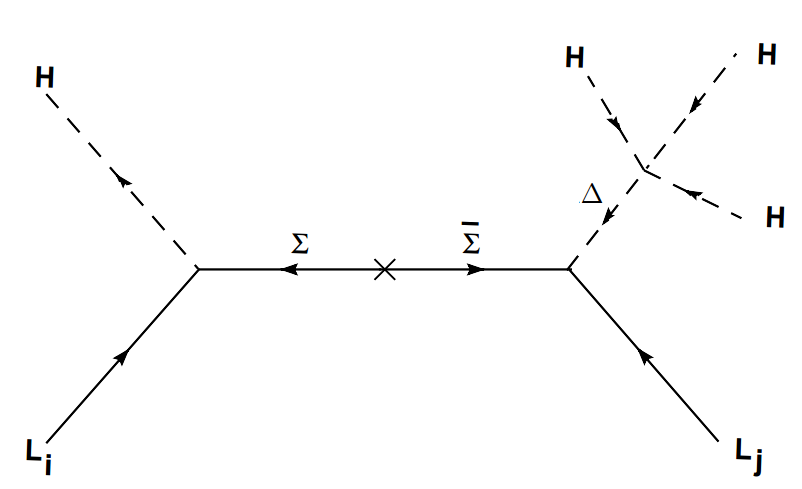}
$$
$$
\includegraphics[height=6.3cm,width=6cm]{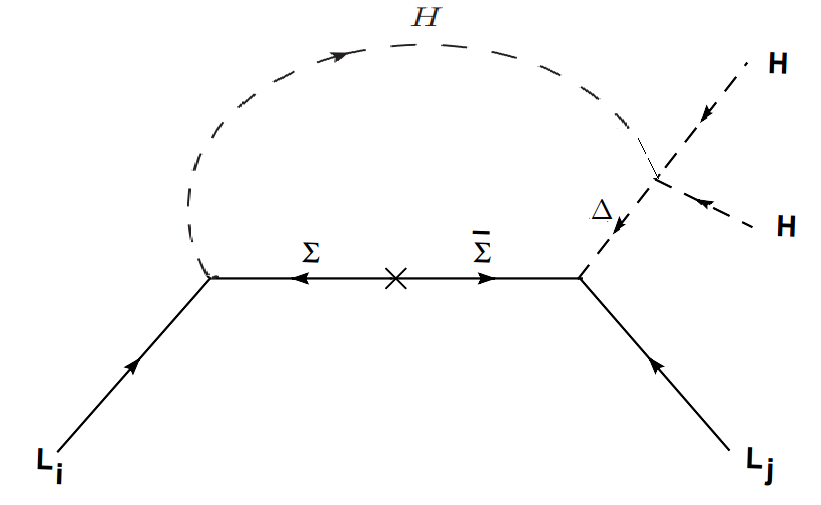}
\includegraphics[height=6.3cm,width=6cm]{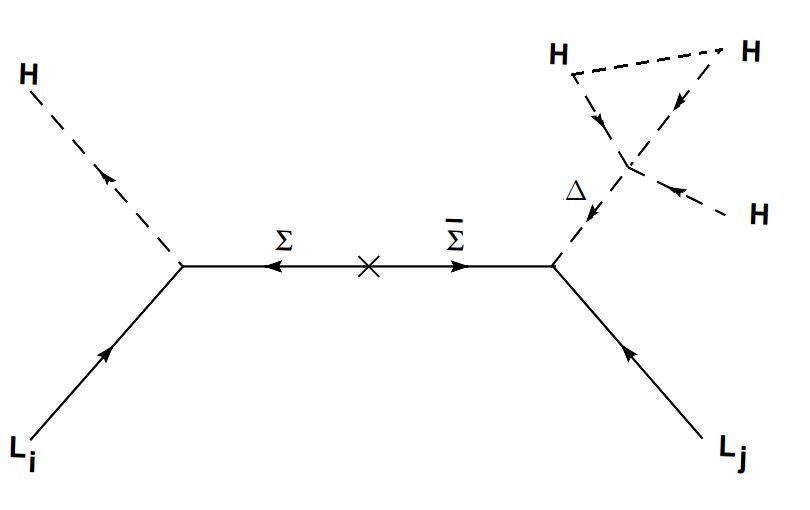}
$$
 \caption{\textbf{Top:} Tree level diagram  generating dimension-7 seesaw operator; \textbf{Bottom:} 1-loop diagrams generating dimension-5 operator for neutrino masses. }
 \label{fig:loop}
 \end{center}
 \end{figure}
 {Substituting the EW VEV , $v_H$, for the Higgs doublet and the induced VEV, $v_\Delta$, for the quadruplet} in Eq.~\ref{neumass}, we obtain  {dimension-7 operator induced} neutrino masses,  $m_{\nu}^{tree}$, as \cite{Babu:2009aq,earlylhc},
\beqa 
(m_{\nu}^{tree})_{ij}  =  
\frac{(Y_i Y_j^{\prime} +  {Y_i^\prime} Y_j) v_\Delta v_{H}}{M_\Sigma} =
-\frac{\lambda_5(Y_i Y_j^\prime +  {Y_i^\prime} Y_j) v_{H}^4}{(M_{\Sigma}
M_{\Delta^0}^2 )}.
\label{eq:mnu2} 
\eeqa
The contribution to the neutrino masses, $m_{\nu}^{loop}$,  {from the loop induced dimension-5 operators} can be computed as \cite{earlylhc},
\begin{equation} 
\label{m_loop}
(m_{\nu})_{ij}^{loop}=\frac{\left(3+\sqrt{3}\right) \lambda _5 v_H^2 M_{\Sigma } 
\left(Y_i Y_j^{'}+ Y_i^{'} Y_j \right)}{16 \pi ^2 \left(M_{\Delta }^2-M_H^2\right)}
\left(
\frac{M_{\Delta}^2 \log \left(\frac{M_{\Sigma }^2}{M_{\Delta }^2}\right)} {M_{\Sigma }^2-M_{\Delta }^2}-
\frac{M_H^2 \log \left( \frac{M_{\Sigma }^2}{M_H^2}\right)} {M_{\Sigma}^2-M_H^2}
\right).
\end{equation}
 {To visualize the relative contribution of the dimension-7 and dimension-5 operators to the neutrino masses, in figure~\ref{fig:contour}, we present a contour plot of the ratio $m_{\nu}^{ loop}/m_{\nu}^{ tree}$ in the $(M_\Delta - M_\Sigma)$ plane. For smaller values of $M_{\Delta}$ and $M_\Sigma$, the dimension-7  (tree level) contribution dominates over dimension-5 (loop level) contribution.}

For completeness of our study, in Table~\ref{numassorder}, we present the  {few benchmark} values of  {$M_\Sigma$,} $v_\Delta$ {, $Y$ and $Y^\prime$} used in our analysis  to generate  {the neutrino masses (presented in the last column of Table~\ref{numassorder}) with} correct order of magnitude.

\begin{table}[htb]\centering
\begin{tabular}{ c c c c c c }
\hline
\hline 
\textbf{Benchmark Point (BP)} &\textbf{ $ M_{\Sigma}$ (TeV)} & \textbf{$ v_{\Delta}$ (GeV)} & \textbf{$Y$} & \textbf{$Y'$} & \textbf{$m_{\nu}$ (eV)}\\
\hline
\hline
\\
\textbf{BP1} & 2 & $10^{-6}$  & $10^{-2}$  & $10^{-2}$ & 0.017  \\
\\
\textbf{BP2} & 3 & $3 \times 10^{-4}$  & $10^{-3}$  & $10^{-3}$ & 0.035  \\
\\
\textbf{BP3} & 4 & $5 \times 10^{-3}$ & $10^{-4}$  & $10^{-3}$ & 0.043  \\
\\
\textbf{BP4} & 2 & $3 \times 10^{-5}$ &  $ 10^{-3} $ & $10^{-2}$ & 0.052  \\ \\
\textbf{BP5} & 3 & $3 \times 10^{-2}$ &  $ 10^{-4} $ & $10^{-4}$ & 0.035  \\ \\
\hline
\hline
\end{tabular} 
\caption{\label{numassorder} Order of neutrino mass for different values of Yukawa couplings $Y$ and $Y^\prime$ for the representative values of  $M_\Sigma$ and $v_\Delta$. Here $v_{H}$ = 174 GeV.}
\end{table} 

\begin{figure}[htb]
\begin{center}
\includegraphics[width=10cm, height=8cm]{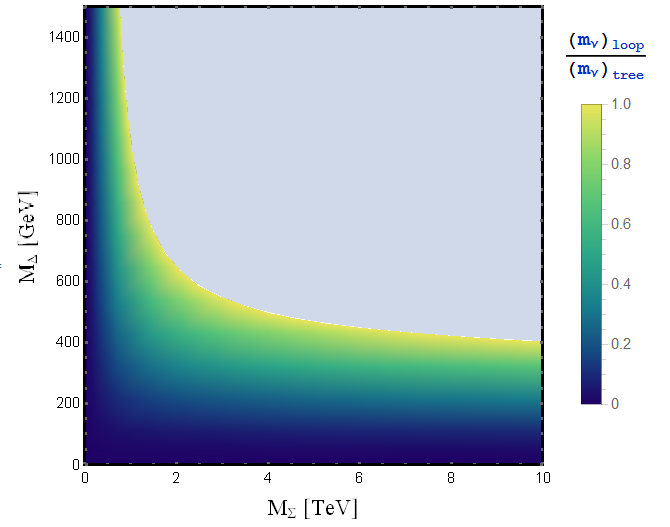}
 \caption { Contour plot of the ratio $m_{\nu}^{loop}/m_{\nu}^{tree}$ 
in the $(M_\Delta - M_\Sigma)$ plane. } 
 \label{fig:contour}
 \end{center}
 \end{figure}

\section{Phenomenology}\label{sec:3}
 {As discussed in the in the previous section, the main motivation for postulation this model is to generate tiny neutrino masses which is achieved by introducing a TeV scale scalar $SU(2)_L$ quadruplet ($\Delta$) and a pair of vector-like $SU(2)_L$ triplet fermions ($\Sigma$ and $\bar \Sigma$). The existence of TeV scale multi-charged scalars (components of $\Delta$) and fermions (components of $\Sigma$ and $\bar \Sigma$) gives rise to the interesting possibility of probing this particular mechanism for neutrino mass generation at the LHC experiment. However, tiny neutrino masses, generated dominantly via tree level effective dimension-7\footnote{The novelty of this model is a new mechanism for generating small neutrino masses which predicts the relation $m_\nu \sim v^4/M^3$, where $v$ is the electroweak scale, rather than the conventional seesaw formula $m_\nu \sim v^2/M$. The new relation $m_\nu \sim v^4/M^3$ arises from the effective dimension-7 operators $LLHH(H^\dagger H)/M^3$ generated at tree level (see Eq. \ref{eq:mnu2}). While the conventional seesaw formula $m_\nu \sim v^2/M$ for neutrino masses via $d = 5$ operators are not induced at tree level, they do arise at 1-loop in this model (see Eq.~\ref{m_loop}). In this study, we are interested in the region of parameter space where the dimension-7 contribution dominates over that coming from the dimension-5 term.} operators, require triplet fermions masses to be at the range of few TeVs (see Fig.~\ref{fig:contour} and Table~\ref{numassorder}).  Therefore, in this work, we have studied the production and signatures of the quadruplet scalars, in particular, multi-charged  quadruplet scalars at the LHC in details. The signatures of the triplet leptons ($\Sigma$ and $\bar \Sigma$) are also discussed briefly in the last part of this section. Being a quadruplet under $SU(2)_L$, the multi-charged scalars can only be pair-produced at the LHC. After being produced in pairs, the quadruplet scalars decays into SM particles giving rise to interesting signatures at the collider. The production and the decay and hence, the resulting collider signatures of the quadruplet scalars are discussed in the following.} 

\subsection{Associated and Pair Production of Charged Higgs}
\begin{figure}[htb]
\begin{center}
\includegraphics[width=12cm, height=3.5cm]{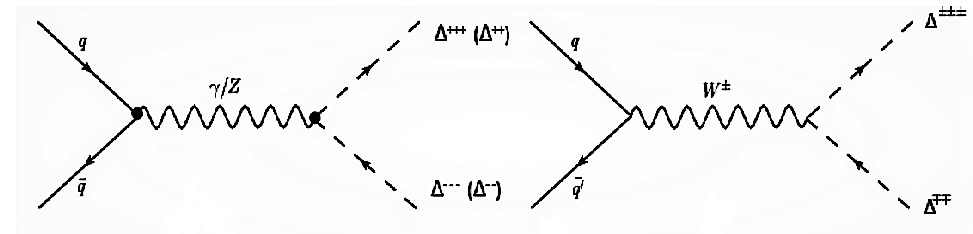}
 \caption {Left : Feynman diagrams for the pair production of $\Delta^{\pm\pm\pm}$ and $\Delta^{\pm\pm}$  via Drell-Yan process. Right : $\Delta^{\pm\pm\pm} \Delta^{\mp\mp}$ are pair produced via s-channel $W^{\pm}$ exchange. } 
 \label{c1}
 \end{center}
 \end{figure}
\begin{figure}[htb]
\begin{center}
\includegraphics[width=13cm, height=10cm]{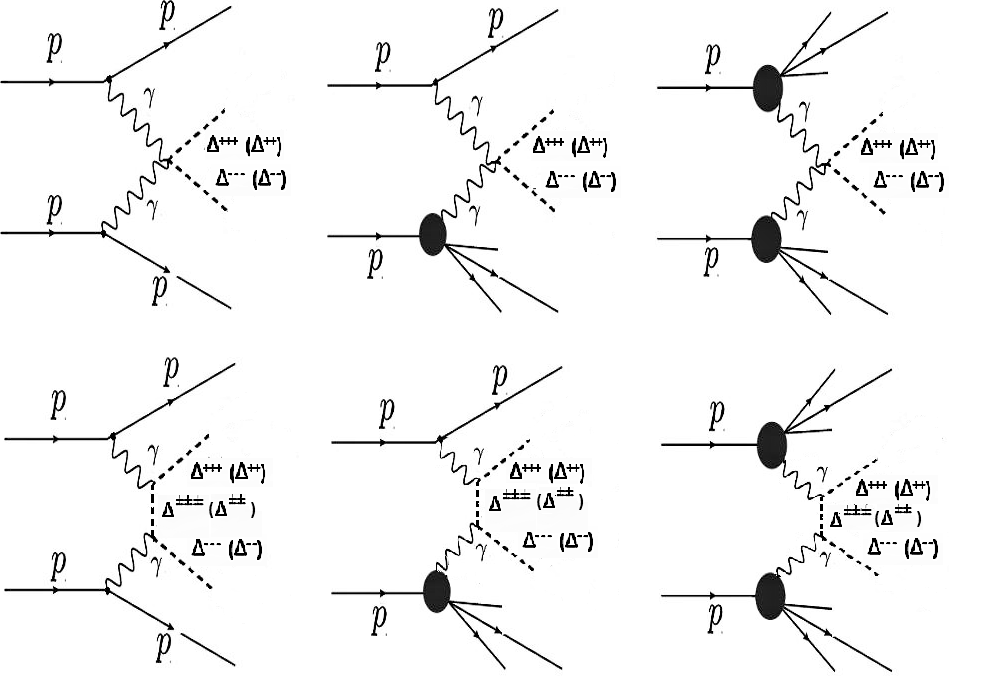}
 \caption {Feynman diagrams for the pair production of $\Delta^{\pm\pm\pm}$ and $\Delta^{\pm\pm}$ via photon-photon fusion processes. Left panel : elastic, middle panel : semi-elastic and Right panel : inelastic scattering sub-processes.} 
 \label{c2}
 \end{center}
 \end{figure}
The LHC  {being a proton-proton collider, the pair productions of} $\Delta^{\pm\pm\pm} \Delta^{\mp\mp\mp}$, $\Delta^{\pm\pm} \Delta^{\mp\mp}$ and $\Delta^{\pm} \Delta^{\mp}$  {take place} via the DY-processes (s-channel $\gamma$ and Z exchanges) [cf. figure \ref{c1}]  {with quark anti-quark in the initial state.}  {Being $s$-channel, Drell Yan pair production cross-sections are significantly suppressed for larger $\Delta^{\pm\pm\pm}/\Delta^{\pm\pm}/\Delta^{\pm\pm}$ masses. However, photo productions of charged scalar pairs ($\gamma \gamma \to \Delta^{\pm\pm\pm} \Delta^{\mp\mp\mp}$, $\Delta^{\pm\pm} \Delta^{\mp\mp}$ and $\Delta^{\pm} \Delta^{\mp}$) take place vis $t(u)$-channel exchange [cf. figure \ref{c2}] of a charged scalar and hence, is not suppressed by the parton center of mass energy. Moreover, the coupling of photon with a pair of charged scalars being proportional to the charge of the scalar, the matrix element squared of photo productions are enhanced by a factor of $3^4$ and $2^4$ for triply and doubly charged scalars, respectively. However, the pair productions of charged scalars at the LHC via photon-photon fusion are suppressed by the small parton density of photon inside a proton.}

 {In fact, the parton density of photon is so small that most of the older versions of PDF's do not include photon as a parton. However, if we want to include QED corrections to the PDF, inclusion of the photon as a parton with an associated parton distribution function is necessary. And in the era of precision physics at the LHC when PDF's are determined upto NNLO in QCD, NLO QED corrections are important (since $\alpha_{s}^2$ is of the same order of magnitude as $\alpha$) for the consistency of calculations. Moreover, as discussed previously, photon-initiated processes could become significant at high energies for some processes. In view of these facts,  NNPDF \cite{Ball:2014uwa,Ball:2013hta}, MRST \cite{Martin:2004dh} and CTEQ \cite{Schmidt:2015zda} have already included photon PDF into their PDF sets. However, different groups used  different approaches for modeling the photon PDF. For example, the MRST \cite{Martin:2004dh} group used a parametrization for the photon PDF based on radiation off of primordial up and down quarks, with the photon radiation cut off at low scales by constituent or current quark masses. The CT14QED \cite{Schmidt:2015zda}  variant of this approach constrains the effective mass scale using $ep \rightarrow e \gamma +X$ data, sensitive to the photon in a limited momentum range through the reaction $e\gamma \rightarrow e \gamma$.  The NNPDF \cite{Ball:2014uwa}\cite{Ball:2013hta} group used a more general photon parametrization, which was then constrained by high-energy W, Z and Drell-Yan data at the LHC.}

 {We have also computed the production of $\Delta^{\pm\pm\pm}$ in association with a $\Delta^{\mp\mp}$. Such a process proceeds through quark anti-quark initial state with the $s$-channel exchange of a $W^\pm$-boson.} The couplings relevant for production and decay of doubly- and triply- charged scalars are shown in Table. \ref{coupt}.
\begin{table}[htb]\centering
\begin{tabular}{ c c }
\hline
\hline 
\textbf{Couplings} &\textbf{Values}\\
\hline
\\
$A^{\mu} \Delta^{\pm\pm\pm} \Delta^{\mp\mp\mp} $& $-3e(p_{1} - p_{2})_{\mu}$\\ 
$A^{\mu} \Delta^{\pm\pm} \Delta^{\mp\mp} $& $-2e(p_{1} - p_{2})_{\mu} $ \\ 
$Z^{\mu} \Delta^{\pm\pm\pm} \Delta^{\mp\mp\mp}$ & $-\dfrac{3e \cos{2 \theta_{w}}}{\sin{2 \theta_{w}}}(p_{1} - p_{2})_{\mu} $\\ 
$Z^{\mu} \Delta^{\pm\pm} \Delta^{\mp\mp}$ & $-\dfrac{2e (\cos{2 \theta_{w}} - 1/2)}{\sin{2 \theta_{w}}}(p_{1} - p_{2})_{\mu} $\\ 
$W^{\mu\mp} \Delta^{\pm\pm\pm} \Delta^{\mp\mp}$ & $\sqrt{3/2} g (p_{1} - p_{2})_{\mu} $ \\ 
$\Delta^{\pm\pm} W^{\mp} W^{\mp}$ & $\sqrt{3}g^{2}v_{\Delta}$\\ 
$\Delta^{\pm\pm} l^{\mp}_{i} l^{\mp}_{j}$ & $\dfrac{m^{\nu}_{ij}}{2\sqrt{3}v_{\Delta}}$\\  
\hline
\hline
\end{tabular} 
\label{cta}
\caption{\label{coupt}The couplings relevant for production and decay of doubly- and triply- charged scalars.}
\end{table}
 {In order to numerically compute the cross-sections,} the model has been implemented in CalcHEP package \cite{Belyaev:2012qa}. For the production cross-sections, we use parton distribution function (PDF) NNPDF23$\char`_$lo$\char`_$as$\char`_$0130 \cite{Ball:2014uwa,Ball:2013hta}, where the photon PDF\footnote{We can also use MRST2004qed$\char`_$proton \cite{Martin:2004dh}, CT14$\char`_$qedinc \cite{Schmidt:2015zda} where the photon PDF  is inclusive, including both inelastic and elastic contributions.} is inclusive with the renormalization and factorization scales being chosen to be the invariant mass of the constituent sub-process. We calculate the pair and associated production cross-sections of $\Delta^{\pm\pm\pm}$ and  $\Delta^{\pm\pm}$ considering both DY and photon-photon fusion processes.  {In figure. \ref{c4}, we have shown the pair and associated production cross-sections of $\Delta^{\pm\pm\pm}$ and $\Delta^{\pm\pm}$ at the 13 TeV LHC considering both DY and photon fusion processes. Figure. \ref{c4} shows that photon fusion significantly contributes to total pair production cross-section of charged scalars for larger masses.}  {For DY process,} the QCD correction  has been also computed, yielding a NLO K-factor of the order of 1.24 at the LHC energy \cite{ATLAS:2016pbt}. But, the noticeable fact is that  {photon-fusion contributes more than the NLO QCD corrections to the DY process for larger masses.} The ratio of the two photon contribution relative to the Drell-Yan channel is shown in figure.\ref{c3}. From the plot (figure.\ref{c3}), we can see that for the higher mass region of $\Delta^{\pm\pm\pm}$ and $\Delta^{\pm\pm}$, photon photon fusion contribution becomes much more significant compared to the DY process. As the pair production cross section is enhanced by $Q^4$, where Q is the charge of the respective charged scalars, the ratio of the two photon contribution relative to the Drell-Yan channel are much more higher  for triply charged Higgs $\Delta^{\pm\pm\pm}$.  {The results of figure.\ref{c3} and figure.\ref{c4} can be summarized as follows:} there is a significant enhancement in  {the total} pair production cross section  {arises from the photon fusion processes and thus},   {photon fusion} can not be ignored   at the LHC  {for the search of multi-charged scalars}, whereas associated production channels remain unaffected.      

\begin{figure}[htb]
\begin{center}
\includegraphics[width=9cm, height=7cm]{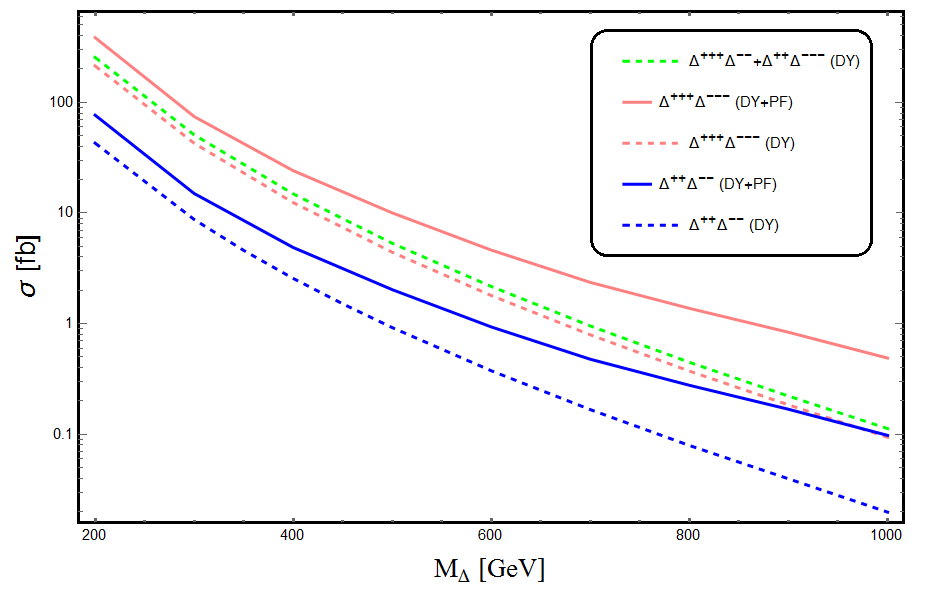}
 \caption {Pair and associated production cross-sections of $\Delta^{\pm\pm\pm}$ and $\Delta^{\pm\pm}$ at the 13 TeV LHC. Red solid (dashed) line is for $\Delta^{\pm\pm\pm}$  pair production cross section via both DY and photon fusion processes (only DY process) and blue solid (dashed) line is for $\Delta^{\pm\pm}$  pair production cross section via both DY and photon fusion processes (only DY process). Green dotted line represents associated production cross section of $\Delta^{\pm\pm\pm}$ and $\Delta^{\pm\pm}$. } 
 \label{c4}
 \end{center}
 \end{figure}

\begin{figure}[htb]
\begin{center}
\includegraphics[width=9cm, height=7cm]{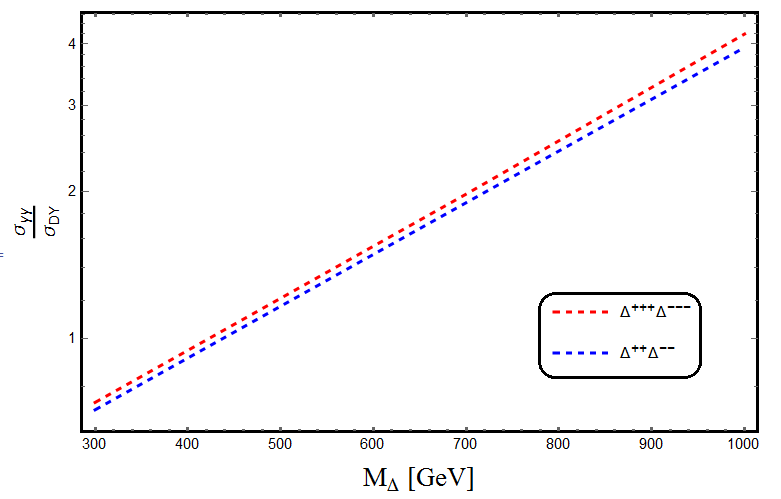}
 \caption {The ratio between $\sigma_{\gamma\gamma}$ and leading order $\sigma_{DY}$ for triply and doubly charged Higgs pair production at the 13 TeV LHC.} 
 \label{c3}
 \end{center}
 \end{figure}

\subsection{Decay Modes of the Charged Higgs}

\begin{figure}[htb]
\begin{center}
\includegraphics[width=15.9cm, height=8.5cm]{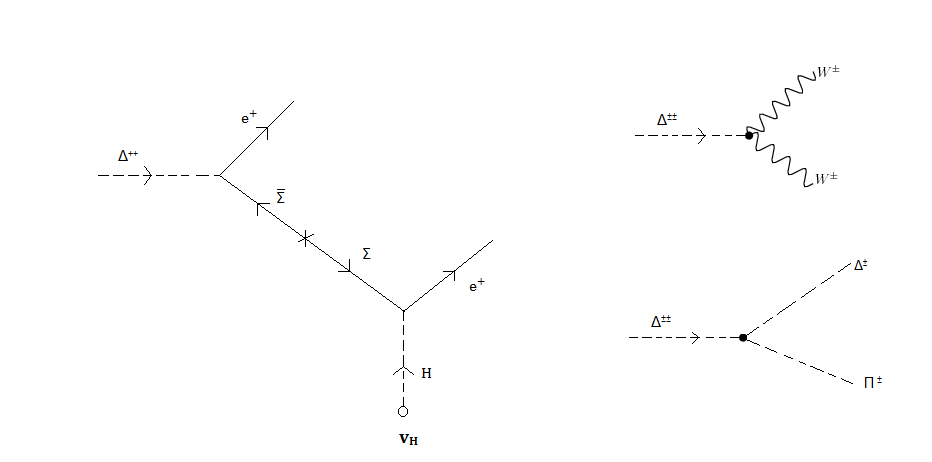}
 \caption {Feynman diagrams for decay of doubly charged scalar $\Delta^{\pm\pm}$ . } 
 \label{c5}
 \end{center}
 \end{figure}

\begin{figure}[htb]
\begin{center}
\includegraphics[width=13cm, height=6cm]{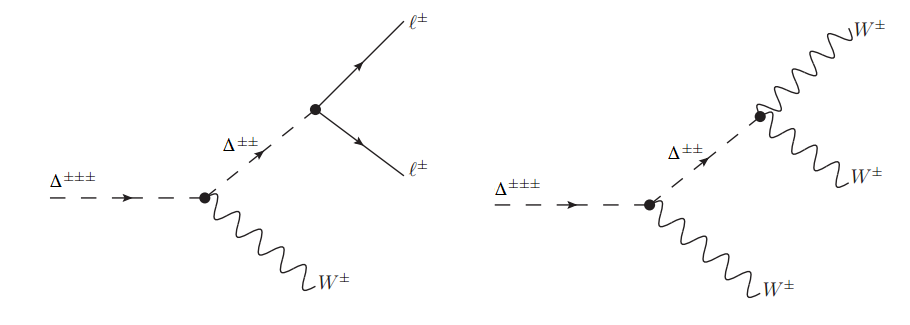}
 \caption {Feynman diagrams for decay of triply charged scalar $\Delta^{\pm\pm\pm}$ . } 
 \label{c6}
 \end{center}
 \end{figure}

{In this section,} we discuss different decay modes of the doubly and triply charged scalars. The representative Feynman diagrams for decays of triply (doubly) charged scalar $\Delta^{\pm\pm\pm}$ ($\Delta^{\pm\pm}$) are shown in figure. \ref{c6} (figure. \ref{c5}). The decay modes of the charged scalars depend on the mass  hierarchy between quadruplet scalars. As noted earlier in section \ref{sec:2}, there are two possible ordering for the masses of the quadruplet scalars depending on the sign of the parameter $\lambda_4$ in the scalar potential. The two possible decay cascades for the triply and doubly charged scalars (depending on the mass hierarchy) are discussed in the following: 

\begin{itemize}
\item Case I : When $\lambda_{4} > 0$, we have $M_{\Delta^{\pm\pm\pm}}< M_{\Delta^{\pm\pm}}< M_{\Delta^{\pm}}< M_{\Delta^{0}}$, so that the triply charged Higgs boson $\Delta^{\pm\pm\pm}$ can only decay to $W^{\pm}l^{\pm}_{i}l^{\pm}_{j}$ or $W^{\pm}W^{\pm}W^{\pm}$. These decays arise through the diagrams where $\Delta^{\pm\pm\pm}$ emits a real $W^{\pm}$ and  an off-shell $\Delta^{\pm\pm}$ which subsequently decays to either two real $W^{\pm}$, or two same sign charged leptons. The corresponding decay rates are given by :

\begin{equation}\label{width}
\Gamma(\Delta^{+++} \rightarrow W^+W^+W^+) ={3 g^6 \over 2048 \pi^3}
{v_\Delta^2 M_\Delta^5 \over m_W^6} I,
\end{equation}

\begin{equation}
\Gamma(\Delta^{+++}
\rightarrow W^+ \ell^+ \ell^+) ={g^2 \over 6144 \pi^3} {M _\Delta
\sum_i m_i^2 \over v_\Delta^2} J,
\end{equation}

where $I,J$ are dimensionless integrals and $m_i$ stands for the light neutrino masses.  In the limit where $M_\Delta \gg m_W$, these integrals are approximately equal to one. 
The partial decay width of $W^{\pm}\ell^{\pm} \ell^{\pm}$ mode scales as  $m_\nu^2/v^2_{\Delta},$ where $m_\nu$ is the light neutrino mass which is proportional to $v_\Delta$. Therefore, the partial decay width $\Gamma(\Delta^{+++}\to W^{\pm}\ell^{\pm} \ell^{\pm})$ is independent of $v_\Delta$. However, $W^{\pm}W^{\pm}W^{\pm}$ mode is proportional to $v^2_{\Delta}$ and hence, the dominant one for larger values of $v_{\Delta}$, while the former is dominant for smaller values of $v_\Delta$.
In figure. \ref{cbr2}, we have shown the variation of branching ratios for the decay modes of triply charged Higgs $\Delta^{\pm\pm\pm}$ as a function of vev $v_{\Delta}$ (left) and mass $M_{\Delta}$ (right). We can see from the plot (see figure. \ref{cbr2}) that when the vev $v_{\Delta}$ is of the order of few KeVs or less,  $\Delta^{\pm\pm\pm}$ dominantly decays to $W^{\pm}\ell^{\pm} \ell^{\pm}$.

\begin{figure}[htb]
\begin{center}
$$
\includegraphics[width=7.5cm, height=7.5cm]{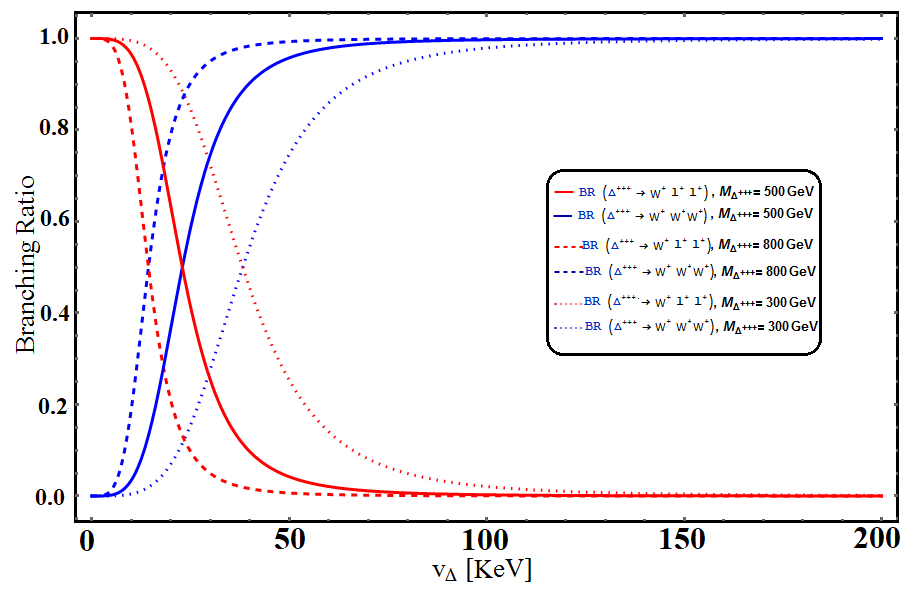}
\includegraphics[width=7.5cm, height=7.5cm]{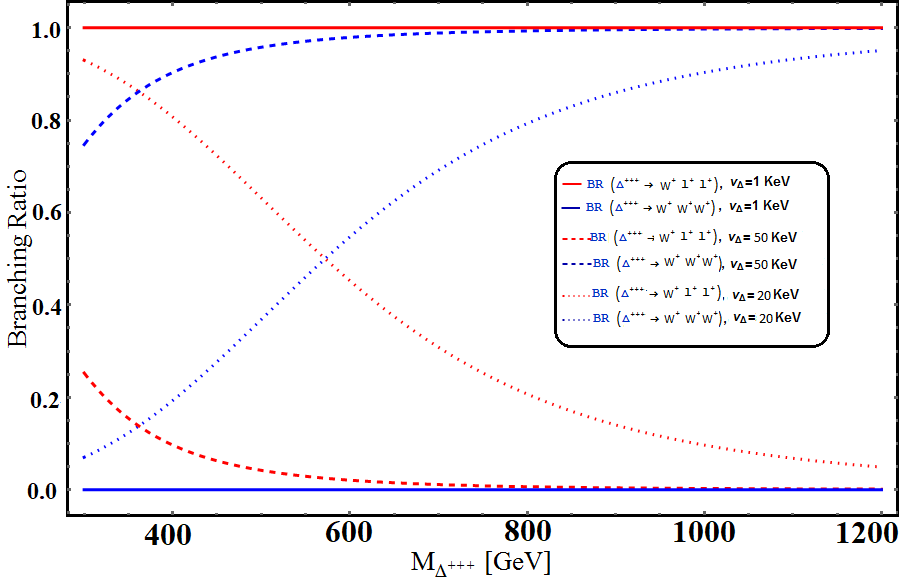}
$$
 \caption { \textbf{Left} : Variation of branching ratio (Br) for different decay modes of $\Delta^{\pm\pm\pm}$ as a function of vev $v_{\Delta}$ for $M_{\Delta^{\pm\pm\pm}}$= 300 (Dotted), 800 (Dashed) and 500 (Solid) GeV. \textbf{Right} : Variation of branching ratio (Br) for different decay modes of $\Delta^{\pm\pm\pm}$ as a function of  mass $M_{\Delta^{\pm\pm}}$ for $v_{\Delta}$= 40 KeV (dotted), 100 KeV (dashed) and 1 KeV (Solid). Red and blue lines are for $W^{+}l^{+}l^{+}$ decay and $W^{+}W^{+}W^{+}$ decay respectively.} 
 \label{cbr2}
 \end{center}
 \end{figure}

The doubly charged Higgs $\Delta^{\pm\pm}$ has the following decay modes: $\Delta^{\pm\pm}\rightarrow W^{\pm}W^{\pm}, \ell^{\pm} \ell^{\pm}$,\ $\Delta^{\pm\pm\pm} W^{\mp*}$ and $\Delta^{+++} \pi^-$. The partial decay widths are given by,
\begin{equation}
\Gamma \left(\Delta^{\pm\pm} \rightarrow l^{\pm}_{i}l^{\pm}_{j} \right) = \dfrac{\mid M_{\nu}^{ij} \mid ^{2}}{8 \pi (1+\delta_{ij})v_{\Delta}^{2}}M_{\Delta^{\pm\pm}},\label{kk1}
\end{equation}

\begin{equation}
\begin{split}
\Gamma \left(\Delta^{\pm\pm} \rightarrow W^{\pm}W^{\pm} \right) = \dfrac{3g^{4} v^{2}_{\Delta}M^{3}_{\Delta^{\pm\pm}}}{32 \pi m^4_{W}}\sqrt{1-\dfrac{4M^{2}_{W}}{M^{2}_{\Delta^{\pm\pm}}}}
\left[1-\dfrac{4M^{2}_{W}}{M^{2}_{\Delta^{\pm\pm}}}+12\dfrac{m_{W}^{4}}{M^4_{\Delta^{\pm\pm}}}\right],
\end{split}
\end{equation}

\begin{equation}
\Gamma \left(\Delta^{\pm\pm} \rightarrow \Delta^{\pm\pm\pm} \pi^{\mp} \right) = \dfrac{3g^4}{32\pi}f_{\pi}^{2}\dfrac{\left(\Delta M \right)^{3}}{m_{W}^{2}},
\end{equation}
where $ M_{\nu}^{ij}$ is the neutrino mass matrix, $\Delta M$ is the mass  splitting between two consecutive members of $\Delta$, $f_{\pi}=130$ MeV, $\delta_{ij}$ is the Kronecker's delta and $l^{\pm}_{i}= e^{\pm},\mu^{\pm}, \tau^{\pm}$. The decay into $\Delta^{\pm\pm\pm} W^{{\star}^{\pm}}$ is suppressed because of the off-shell $W^{\pm}$-boson ($W^{\pm *}$) in the final state. We note that the decay width for the decay mode $\Delta^{\pm\pm} \rightarrow l^{\pm}l^{\pm}$ is
independent of $v_\Delta$ (in Eq.~\ref{kk1}, the $v_\Delta$ dependence in $\mid M_{\nu}^{ij} \mid ^{2}$ in the numerator cancels with the $v_\Delta^2$ in the denominator), the decay width to $W^{\pm}W^{\pm}$ final state is proportional to $v_{\Delta}^{2}$, while the one to $\Delta^{\pm}\pi^{\pm}$ is independent of $v_{\Delta}$, and proportional to $(\Delta M)^{3}$. In  figure.~\ref{cbr1}, we plot the relative branching ratios of $\Delta^{\pm\pm}$ as a function of $M_{\Delta}$ (right) and $v_{\Delta}$ (left). For simplicity, we have taken the masses of the quadruplets to be the same\footnote{Constraints from the $\rho$ parameter dictates the splitting to be $< 38 ~{\rm GeV}$ \cite{Babu:2009aq, earlylhc}, and can be even smaller depending on the values of $\lambda_4$.}. As expected, for a very small $v_{\Delta}$ ($\lesssim
10^{-4}$ GeV), the decay to $l^{\pm}l^{\pm}$ dominate, whereas for higher values of $v_{\Delta}$ ($\gtrsim 10^{-4}$ GeV), the mode $W^\pm W^\pm$ dominate. For completeness, we have also done the calculation for a small mass splitting of 2.5 GeV and we get that for the vev $v_{\Delta} \leq 1.5$ KeV the branching ratio to same sign dilepton becomes 100$\%$. The branching ratio study for different decay modes of $\Delta^{\pm\pm}$ for non-degenerate masses of $\Delta$ members can be found in our earlier paper \cite{Babu:2009aq,earlylhc2}.
\begin{figure}[htb]
\begin{center}
$$
\includegraphics[width=7.5cm, height=7.5cm]{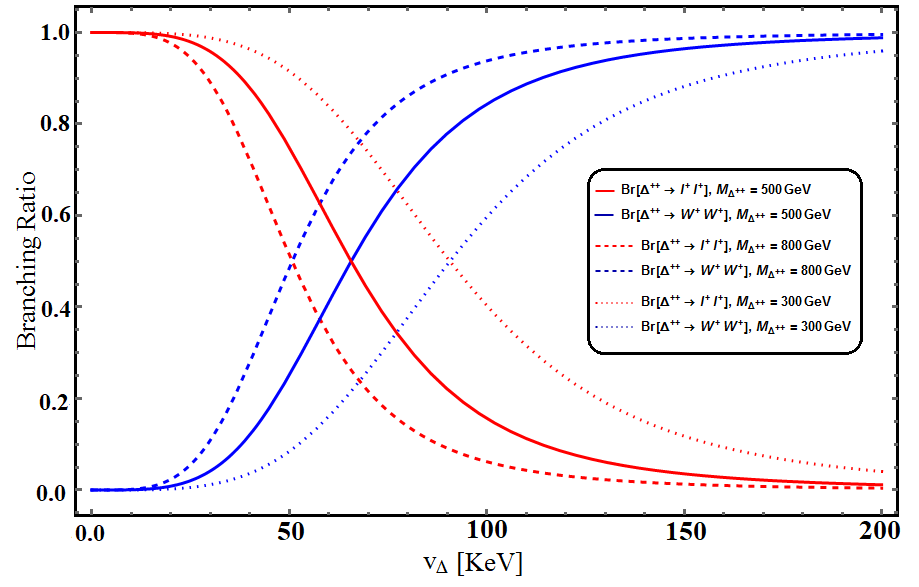}
\includegraphics[width=7.5cm, height=7.5cm]{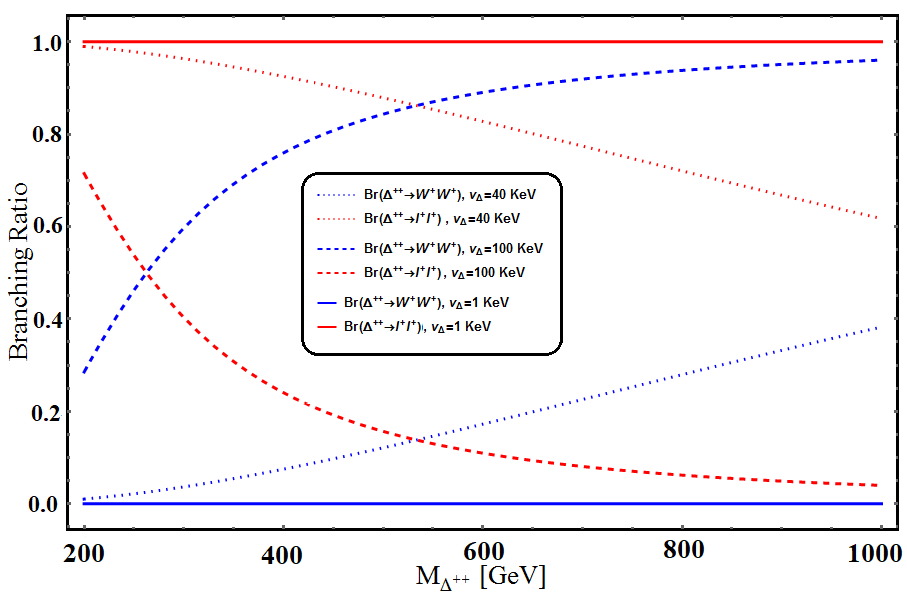}
$$
 \caption { \textbf{Left} : Variation of branching ratio (Br) for different decay modes of $\Delta^{\pm\pm}$ as a function of vev $v_{\Delta}$ for $M_{\Delta^{\pm\pm}}$= 300 (Dotted), 800 (Dashed) and 500 (Solid) GeV. \textbf{Right} : Variation of branching ratio (Br) for different decay modes of $\Delta^{\pm\pm}$ as a function of  mass $M_{\Delta^{\pm\pm}}$ for $v_{\Delta}$= 40 KeV (dotted), 100 KeV (dashed) and 1 KeV (Solid). Red and blue lines are for same sign dilepton decay and same sign diboson decay respectively.} 
 \label{cbr1}
 \end{center}
 \end{figure}

\item Case II :  When $\lambda_{4}< 0$, we have $M_{\Delta^{\pm\pm\pm}}> M_{\Delta^{\pm\pm}}> M_{\Delta^{\pm}}> M_{\Delta^{0}}$. If the quadruplet components are not degenerate and $\Delta^{\pm\pm\pm}$ is the heaviest member in the quadruplet, then $\Delta^{\pm\pm\pm}$ decays to $\Delta^0$ and SM particles via cascades involving other quadruplet scalars: $\Delta^{\pm\pm\pm} \rightarrow \Delta^{\pm\pm}W^{*\pm} \rightarrow W^{*\pm}W^{*\pm}\Delta^{\pm}\rightarrow W^{*\pm}W^{*\pm}W^{*\pm}\Delta^{0}$. The other possible decay mode of $\Delta^{\pm\pm\pm}$ is into a $\Delta^{\pm\pm}$ in association with a $\pi^{\pm}$. For large enough mass splitting between the quadruplet scalars, cascade decay dominates over the decay into $\Delta^{\pm\pm} \pi^{\pm}$.
\end{itemize}

\subsection{Collider Phenomenology}
\begin{figure}[b!]
\begin{center}
\includegraphics[width=7cm, height=7cm]{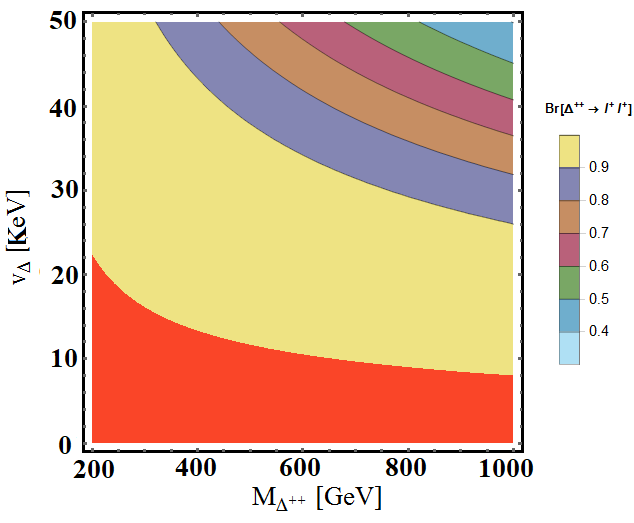}
\includegraphics[width=7cm, height=7cm]{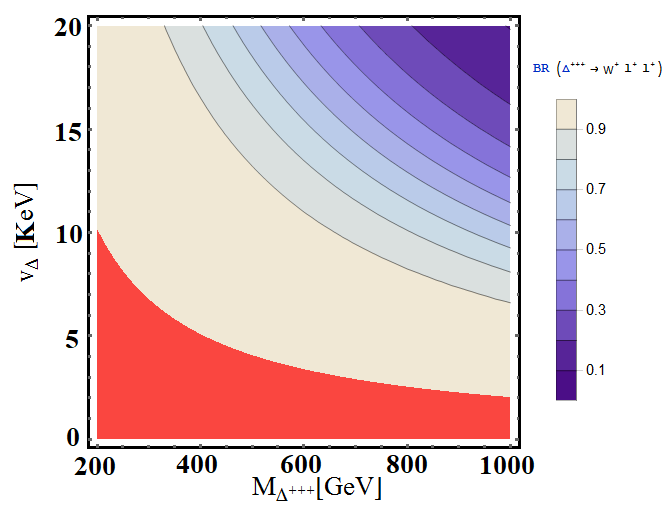}
 \caption {Contour plot for branching ratio Br$\left(\Delta^{\pm\pm} \rightarrow l^{\pm}l^{\pm}\right)$ (left panel) and Br$\left(\Delta^{\pm\pm\pm} \rightarrow W^{\pm}l^{\pm}l^{\pm}\right)$ (right panel) in $v_{\Delta}$-$M_{\Delta^{\pm\pm}}$ plane. Branching ratio scale is shown in right side of the figure. Red shaded zone in both figure corresponds to Br$\left(\Delta^{\pm\pm} \rightarrow l^{\pm}l^{\pm}\right)$ or Br$\left(\Delta^{\pm\pm\pm} \rightarrow W^{\pm}l^{\pm}l^{\pm}\right)$ $\sim 100 \%$.} 
 \label{con1}
 \end{center}
 \end{figure}
In this section, we studied the collider signatures of multi-charged scalars for positive\footnote{For negative $\lambda_4$, $\Delta^{+++}$ being the heaviest among the $\Delta$'s, decays to $\Delta^{++}$ in association with a off-shell $W$-boson. $\Delta^{++}$ subsequently decays to $\Delta^+$ followed by the $\Delta^+$ decay to $\Delta^0$. Therefore, for negative $\lambda_4$, the decay of $\Delta^{+++}$ also gives rise to 3 same charge leptons (when the off-shell $W$'s decays leptonically). However, these leptons are very soft because of small mass-splitting between the components of the quadruplet scalar and mostly fall outside detector acceptance.} $\lambda_4$ where $\Delta^{+++}$ is the lightest among the $\Delta$'s. First, we focus on the same-sign dilepton decay mode of $\Delta^{\pm\pm}$. The same-sign dilepton decay of $\Delta^{\pm\pm} \to l^\pm l^\pm$ is characterized by a invariant mass peak at $m_{\Delta^{\pm\pm}}$ in the same-sign dilepton invariant mass distribution. In view of negligible\footnote{Same-sign dilepton in the SM arises from the multiple $W^\pm$ and $Z$-bosons production which are quite suppressed. For example, SSD can arise from 3 $W^\pm$-bosons ($pp\to W^\pm W^\pm W^\mp$) production followed by leptonic decay of 2 same-sign $W^\pm$-bosons and hadronic decay of the other. $W^\pm Z$ pair production also contributes to the background when both $W^\pm$ and $Z$-boson decays leptonically and one lepton from $Z$-decay falls out side detector coverage. Semi-leptonic decay of $t\bar t$ pairs also contributes to the SSD background when one $b$-quark decays leptonically. Though the leptons from the $b$-decay are usually rejected by the lepton isolation criteria, a non-negligible SSD background arises from $t\bar t$ production due to its huge cross-section at the LHC. Miss identification of a jet as lepton and charge miss-measurement of leptons also contributes to the background. However, all these backgrounds are estimated to be small. Moreover, the background same-sign dileptons are not characterized by any invariant mass peak.} SM background, same-sign dilepton channel characterized by an invariant mass peak in the dilepton invariant mass distribution is considered to be one of the cleanest channel to search at the LHC. Since we are interested mostly on the like-sign dilepton decay of $\Delta^{\pm\pm}$ and the LHC has already searched for a invariant mass peak in the like-sign dilepton invariant mass distribution, it is important to pin down the part of parameter space for which  $\Delta^{\pm\pm}$ dominantly decays to dileptons. In figure \ref{con1} (left panel), we have shown the contour plot for branching ratio Br$\left(\Delta^{\pm\pm} \rightarrow l^{\pm}l^{\pm}\right)$ in $v_{\Delta}$-$M_{\Delta^{\pm\pm}}$ plane. Figure \ref{con1} (left panel) shows that for low $v_{\Delta}$, $\Delta^{\pm\pm}$ dominantly decays to dileptons. Therefore, it is possible to exclude low $v_\Delta$ region of this model from the LHC bounds on the cross-section of the resonant same-sign dilepton production. The exclusion limits in the context of this model will be discussed in the next section.

Other characteristic feature of this model is the existence of a triply charged scalar. The pair production cross-section of triply charged scalar is relatively large (see Figure \ref{c4}) at the LHC because of its enhanced coupling with photon. After being produced, triply charged scalars decays into $W^\pm W^\pm W^\pm$ or $W^\pm l^\pm l^\pm$ depending on the part of parameter space. In figure. \ref{con1}, we have shown the contour plot for branching ratio Br$\left(\Delta^{\pm\pm\pm} \rightarrow W^{\pm}l^{\pm}l^{\pm}\right)$ in $v_{\Delta}$-$M_{\Delta^{\pm\pm\pm}}$ plane. Figure. \ref{con1} shows that for low $v_\Delta$,  $\Delta^{\pm\pm\pm} \to W^{\pm}l^{\pm}l^{\pm}$ decay dominates over $\Delta^{\pm\pm\pm} \to W^{\pm} W^{\pm} W^{\pm}$. In both cases, the pair production and decay of $\Delta^{\pm\pm\pm}$ give rise to interesting multi-lepton (6,5,4-leptons, same-sign three leptons e.t.c.) final states which will be discussed in the subsequent sections. 

The pair production cross-sections of the doubly and triply charged scalars are completely determined by the scalar masses and the SM gauge couplings. The subsequent decay branching ratios of these scalars depend on the induced VEV, $v_\Delta$. Therefore, the collider signatures of the charged scalars crucially depend on $v_\Delta$. For example, the pair-production of doubly charged scalar give rise to two pairs of resonant same-sign dilepton for $v_\Delta<v^c_\Delta$ where $v_\Delta^c$ is the critical value of $v_\Delta$ below which $\Delta^{\pm\pm}$ dominantly decays to dileptons. For the decay modes of $\Delta^{\pm\pm}$, the critical value of $v_\Delta$ is given by $v_\Delta^c \sim 0.5 \sqrt{m_\nu/M_\Delta} v_H$. For few hundred GeV $\Delta^{\pm\pm}$ masses ($m_\nu \sim 0.1$ eV and $v_H \sim 250$ GeV), $v_\Delta^c$ is estimated to be in the range of few tens of KeV to hundred KeV. On the other hand, for the decay modes of $\Delta^{\pm\pm\pm}$, the critical value of $v_\Delta$ is given by $v_\Delta^c \sim 0.5 M_\Delta^{-1} \sqrt{M_W m_\nu} v_H$. In this case, $M_\Delta \sim$ few hundred GeV corresponds to a $v_\Delta^c$ in the range of few KeV to few tens of KeV. For both $\Delta^{\pm\pm}$ and $\Delta^{\pm\pm\pm}$, the critical value of the $v_\Delta$ is essentially determined by $M_\Delta$. For $M_\Delta$ of few hundred GeV, the critical value of $v_\Delta$ varies between few KeV to hundred KeV. Therefore, for the collider analysis, we have defined two $v_\Delta$ regions namely, the "Large $v_\Delta$" region and the "Small $v_\Delta$" region as $v_\Delta> 100$ KeV and $v_\Delta<1$ KeV, respectively. In the "Large $v_\Delta$" region, the multi-charged scalars dominantly decay into $W^\pm$-bosons and in the "Small $v_\Delta$" region, they decay into same-sign leptons.

\subsubsection{Bound on Doubly Charged scalar}
\begin{figure}[htb]
\begin{center}

\includegraphics[width=10cm, height=8cm]{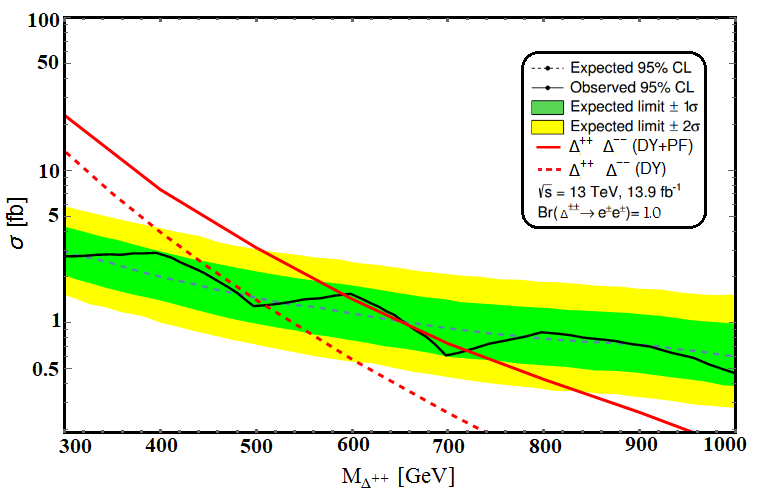}

 \caption {The observed and expected 95$\%$ C.L. upper limits of the production cross-section times branching ratio to electrons \big[$\sigma(\Delta^{++}\Delta^{--})\times Br(\Delta^{\pm \pm}\to e^\pm e^\pm)$\big] as a function of $M_{\Delta^{\pm\pm}}$ using ATLAS results \cite{ATLAS:2016pbt} at $\sqrt{s}$ = 13 TeV with 13.9 fb$^{-1}$ integrated luminosity. The theoretical prediction for $\sigma(\Delta^{++}\Delta^{--})\times Br(\Delta^{\pm \pm} \to e^\pm e^\pm)$ in the context of present model for a $SU(2)_L$ quadruplet doubly charged scalar are presented by red solid (photon fusion + DY) and dashed (DY-only) lines. In the calculation of the theoretical cross-section, we have assumed $Br(\Delta^{\pm \pm} \to e^\pm e^\pm) \sim 100\%$.} 
 \label{figbound}
 \end{center}
 \end{figure}

In the context of LR-symmetry\footnote{ In case of Minimal Left Right Symmetric model (MLRSM), $\Delta^{\pm \pm}$ is a part of SU(2)$_L$ triplet, whereas $\Delta^{\pm \pm}$  is a part of SU(2)$_L$ quadruplet in our model and hence, $Z \Delta^{\pm \pm} \Delta^{\mp \mp} $ coupling is completely different for both the cases and it largely effects Drell Yan pair production through s-channel Z exchange at the LHC. In MLRSM, there is no triply charged scalar, whereas in our model, the scalar SU(2)$_L$ quadruplet contains a triply charged scalar field $\Delta^{\pm \pm \pm}$. As a consequence, our model predicts same sign trilepton events in addition to the same sign dilepton signatures at the collider experiments. Same sign trilepton events are unique in this model compared to the MLRSM. The collider signature of triply charged scalars will be discussed in a later section.\color{black}}, the ATLAS Collaboration has recently searched \cite{ATLAS:2016pbt} for the doubly-charged scalar decaying into a pair of like-sign leptons in the same-sign di-electrons invariant mass spectrum with luminosity 13.9 $fb^{-1}$ at $\sqrt{S}$ = 13 TeV. In absence of any significant deviation from the SM background prediction, limits are imposed on the doubly charged scalar pair production cross-section times branching ratio to leptons ($\sigma(\Delta^{++}\Delta^{--})\times Br(\Delta^{\pm \pm} \to l^\pm l^\pm)$). In the context of LR-symmetric model, the bound on the $\sigma(\Delta^{++}\Delta^{--})\times Br(\Delta^{\pm \pm} \to l^\pm l^\pm)$ corresponds to a lower limit of 570 (420) GeV on the mass of  doubly charged $SU(2)_L$ triplet(singlet) scalar assuming its 100$\%$ branching ratio to a pair of same-sign electrons.

In the context of our model, the pair production and subsequent leptonic decay of the doubly charged scalar ($\Delta^{\pm\pm}$) gives rise to similar signature at the LHC and hence, our model also should comply with non-observation of any excess in same sign dilepton search. As a result, the model independent limits on $\sigma(\Delta^{++}\Delta^{--})\times Br(\Delta^{\pm \pm}\to l^\pm l^\pm)$ is also applicable in our model where the doubly charged scalars are quadruplet under $SU(2)_L$. In Figure \ref{figbound}, we compare theoretical pair production cross-sections of doubly charged quadruplet scalars with 13 TeV ATLAS limit \cite{ATLAS:2016pbt} on $\sigma(\Delta^{++}\Delta^{--})\times Br(\Delta^{\pm \pm}\to l^\pm l^\pm)$. The solid black line in Figure \ref{figbound} corresponds to 95$\%$ C.L. on upper limit on $\sigma(\Delta^{++}\Delta^{--})\times Br(\Delta^{\pm \pm}\to l^\pm l^\pm)$ obtained by ATLAS collaboration with 13 TeV center of mass energy and 13.9 fb$^{-1}$ integrated luminosity. The green and yellow bands correspond to the 1$\sigma$ and 2$\sigma$ bands on the expected limits respectively. As discussed in Section 3.1, the photon fusion contributes significantly to total production cross-section of multi-charged scalar pairs. Therefore, irrespective of origin of $\Delta^{\pm\pm}$, one must incorporate photon fusion contribution to the total pair production cross-section in addition to DY-contribution. However, Ref. \cite{ATLAS:2016pbt} considered only DY-production of $\Delta^{\pm\pm}$ pairs in the context of LR-symmetry and hence, significantly under estimated the mass limits on the doubly charged scalars in LR-symmetry \cite{Babusudip}. In order to quantify the effect of photon fusion contribution on the bound of $\Delta^{\pm\pm}$ mass, in figure \ref{figbound}, we have presented the theoretical values for $\sigma(\Delta^{++}\Delta^{--})\times Br(\Delta^{\pm \pm}\to l^\pm l^\pm)$ in the context of a doubly charged $SU(2)_L$ quadruplet assuming  $Br(\Delta^{\pm \pm} \to e^\pm e^\pm)\sim 100\%$ for DY-production only (red dashed line) as well as DY plus photon fusion (red solid line). Figure \ref{figbound} shows that as a result of including photon fusion contribution, there is a significant enhancement on the lower bound of $\Delta^{\pm\pm}$ mass. A brief summary of the 95$\%$ CL exclusion limits on $M_{\Delta^{\pm\pm}}$ using ATLAS preliminary results at $\sqrt{s}$ = 13 TeV with 13.9 fb$^{-1}$ integrated luminosity is shown in Table \ref{bound}. It is important to note that there are some uncertainties in photon PDF \cite{Ball:2014uwa, Ball:2013hta,Martin:2004dh,Schmidt:2015zda} selection. We estimated that the uncertainty in photon PDF selection corresponds to a uncertainty $\pm$13 GeV on $M_{\Delta^{\pm\pm}}$ limits.
\begin{table}[htb!]
\centering
  \begin{tabular}{c c c c c }
   \hline
   \hline
    \multirow{2}{*}{\textbf{Benchmark Point}} &
      \multicolumn{3}{p{4.3cm}}{\textbf{Limits on $M_{\Delta^{\pm\pm}}$ (GeV)}} \\
       &   \textbf{(DY)} &\textbf{(DY+PF)}
      \\
    \hline
    \\
    \textbf{$\Delta^{\pm\pm} \rightarrow e^{\pm}e^{\pm}$= 100$\%$   }& \ 509 & \ \ \textbf{725} \\ \\
    \textbf{$\Delta^{\pm\pm} \rightarrow e^{\pm}e^{\pm}$= 50$\%$   }& \ 368 & \ \ \textbf{521}\\ \\
 \textbf{   $\Delta^{\pm\pm} \rightarrow e^{\pm}e^{\pm}$= 33$\%$ }  & \ 330 & \ \ \textbf{387} \\ \\
    \hline
    \hline
  \end{tabular}
  \caption{Summary of the 95$\%$ CL exclusion limits on $M_{\Delta^{\pm\pm}}$ using ATLAS results at $\sqrt{s}$ = 13 TeV with 13.9 fb$^{-1}$ integrated luminosity. DY : Drell-Yan pair production; PF : photon fusion process. }
  \label{bound}
\end{table}

\begin{figure}[htb]
\begin{center}

\includegraphics[width=12cm, height=8cm]{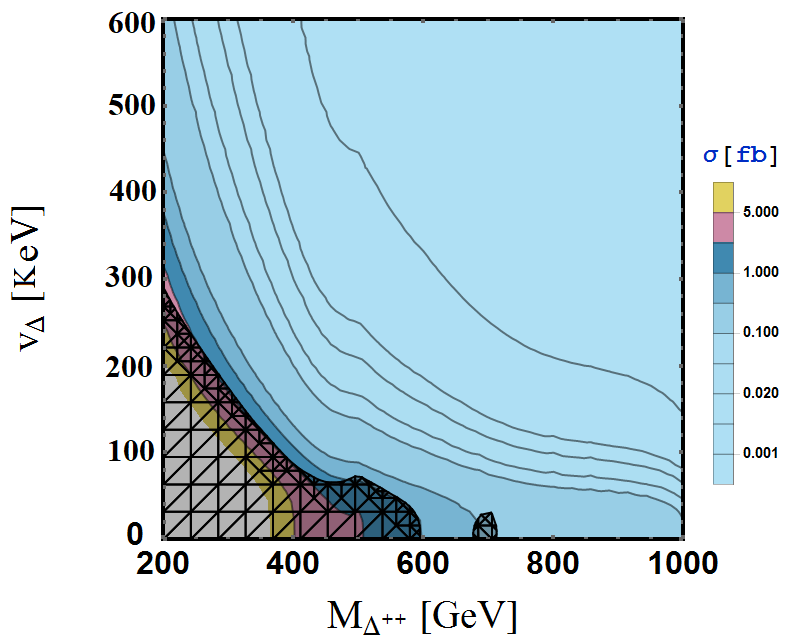}

 \caption {Contour plot of $\sigma(\Delta^{++}\Delta^{--})\times Br(\Delta^\pm\to e^\pm e^\pm)$ on $v_{\Delta}$-$M_{\Delta^{\pm\pm}}$ plane. The crossed region of the plot is excluded from the ATLAS search \cite{ATLAS:2016pbt} for same sign dilepton invariant mass peak at 13 TeV center of mass energy and 13.9 fb$^{-1}$ integrated luminosity.} 
 \label{exclusion}
 \end{center}
 \end{figure}

The production cross-section of a pair of doubly charged $SU(2)_L$ quadruplet scalars at the LHC is completely determined by the mass of $\Delta^{\pm\pm}$. On the other hand, as discussed in details in Section 3.2, the decay branching ratio of $\Delta^{\pm\pm}$ into a pair of leptons is mainly determined by the induced VEV $v_\Delta$. Therefore, the ATLAS upper bound on $\sigma(\Delta^{++}\Delta^{--})\times Br(\Delta^{\pm \pm} \to e^\pm e^\pm)$ in Figure \ref{figbound} can be used to exclude parts $v_{\Delta}$-$M_{\Delta^{\pm\pm}}$ plane. In Figure \ref{exclusion}, we present contour plot of $\sigma(\Delta^{++}\Delta^{--})\times Br(\Delta^{\pm \pm}\to e^\pm e^\pm)$ on $v_{\Delta}$-$M_{\Delta^{\pm\pm}}$ plane. The crossed region of Figure \ref{exclusion} is excluded from the ATLAS search \cite{ATLAS:2016pbt} for same sign dilepton invariant mass peak at 13 TeV center of mass energy and 13.9 fb$^{-1}$ integrated luminosity.

\subsubsection{Characteristic signatures of (multi-)charged scalars at the LHC}
\begin{figure}[htb]
\begin{center}
\includegraphics[width=12cm, height=8cm]{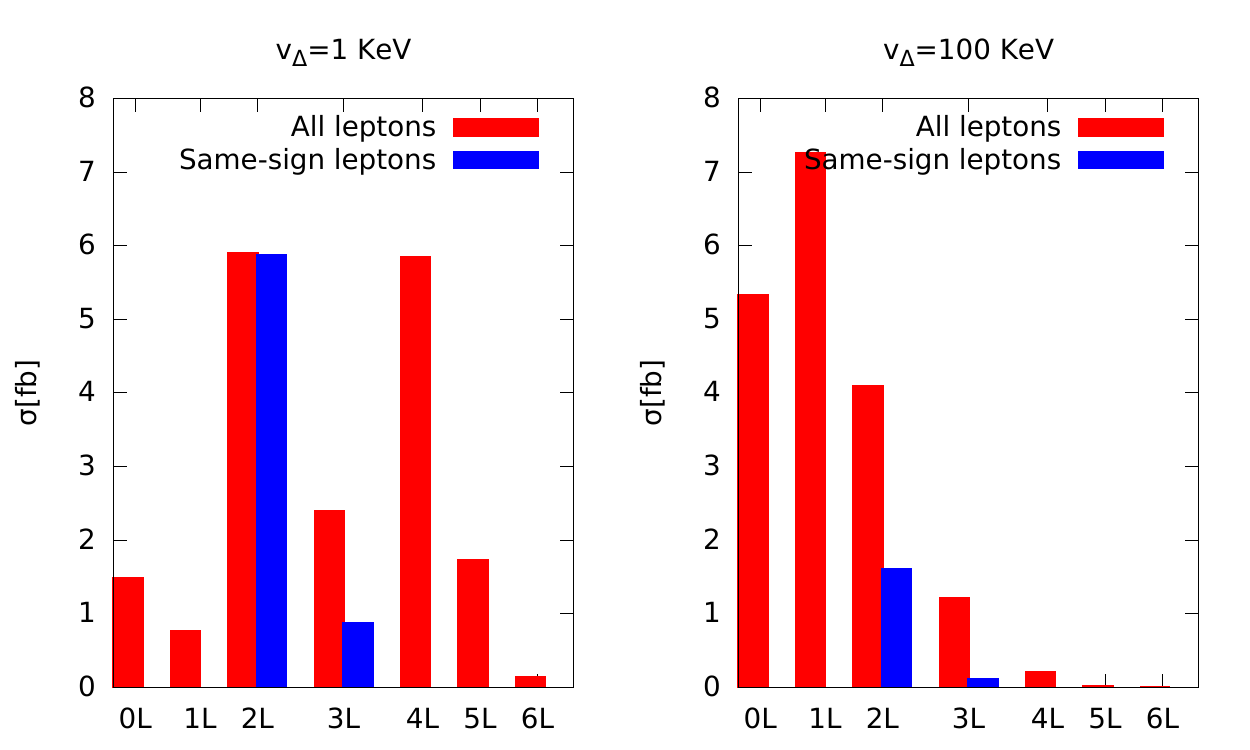}
\caption {Lepton multiplicity distribution after imposing the acceptance cuts summarized  in  Eqs. (\ref{cut:pT}--\ref{cut:jj-iso}). We have considered $m_\Delta=500$ GeV. Left panel corresponds to small $v_\Delta$ and right panel corresponds to large $v_\Delta$.} 
\label{nlep}
\end{center}
\end{figure}

After discussing the production, decay and collider bounds on the quadruplet scalars, we are now equipped enough to discuss the characteristic collider signatures of these scalars at the future runs of the LHC. As discussed in the previous section, for small $v_\Delta$, multi-charged quadruplet scalars dominantly decay into leptonic final states and hence, give rise to lepton rich signatures at the LHC. On the other hand, for large $v_\Delta$, quadruplet scalars dominantly decay to $W$-bosons and subsequent leptonic decay of $W$-bosons give rise to leptons in the final state. Though, the leptonic final states for large $v_\Delta$ are suppressed by the leptonic branching ratio of the $W$-boson, muti-leptons signatures are considered very promising because of small or negligible SM background. In this work, we have studied multi-leptons signatures of the charged-quadruplet scalars. Since the detection efficiencies of electrons and muons are much higher than the taus, for the rest of this work, we have only considered electrons and muons as leptons. Pair and associated production of doubly and triply charged scalars give rise to final states with 0--6 leptons multiplicity including interesting same-sign dileptons (SSD) and same-sign 3-leptons (SS3L) events. However, before going into the discussion of lepton multiplicity as well as other characteristic kinematic distributions, it is important to list a set of basic requirements for leptons and jets to be visible at the detector.

\begin{figure}[htb]
\begin{center}
\includegraphics[width=12cm, height=12cm]{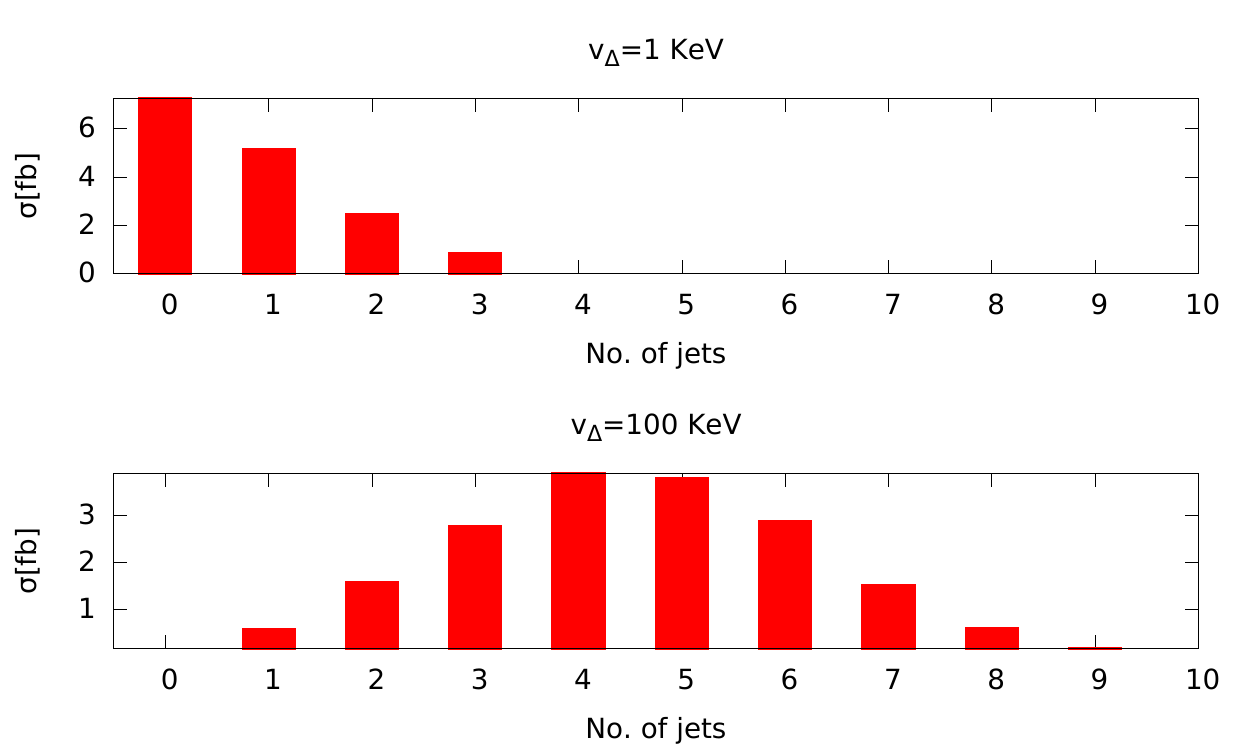}
\caption {Jet multiplicity distribution after imposing the acceptance cuts summarized  in  Eqs. (\ref{cut:pT}--\ref{cut:jj-iso}). We have considered $m_\Delta=500$ GeV. Top panel corresponds to small $v_\Delta$ and bottom panel corresponds to large $v_\Delta$.} 
\label{njet}
\end{center}
\end{figure}

It should be noted that any realistic detector has only a finite resolution; this applies to both energy/transverse momentum measurements as well as the determination of the angle of motion. For our purpose, the latter effect can be safely neglected\footnote{The angular resolution is, generically, far
superior to the energy/momentum resolutions and too fine to be of any
consequence at the level of sophistication of this analysis.} 
and we simulate the former by smearing the energy with Gaussian
functions. The energy resolution function receives contributions from
many sources and are, in general, a function of the detector
coordinates. We, though, choose to simplify the task by assuming a
flat resolution function equating it to the worst applicable for our
range of interest \cite{gsmear}, namely, 
\be
\frac{\Delta E}{E}=\frac{a}{\sqrt {E/{\rm GeV}}}\oplus b,
\ee
where,  $ a=100\%, b=5\%$ for jets and $a=15\%$ and $b=1\%$ for leptons, and $\oplus $ denotes a sum in quadrature. Keeping in mind the LHC environment as well as the detector configurations, we demand that, to be visible, a lepton or jet must have an adequately large transverse momentum and they are well inside 
the rapidity coverage of the detector, namely,
\be
p_T^{l} > 20~GeV \ ;~~~~~~p_T^{j} > 20~GeV \ ,
\label{cut:pT}
\ee
\be
|\eta_{l}| \leq 2.5 \ ;~~~~~~ |\eta_{j}| \leq 2.5 \ .
\label{cut:eta}
\ee
We demand that a lepton be well separated from other leptons and jets so that they can be identified as individual physics objects. We use the well-known cone algorithm defined in terms of a cone angle $\Delta R_{ij} \equiv \sqrt{\left (\Delta \phi_{ij}\right)^2 + \left(\Delta \eta_{ij}\right)^2} $, with 
$\Delta \phi $ and $ \Delta \eta $ being the azimuthal angular 
separation and rapidity difference between two particles.
Quantitatively, we impose
\be
\Delta R_{l \, l} > 0.4;~~~~~~
\Delta R_{l \, j} > 0.4;~~~~~~
\Delta R_{j \, j} > 0.7.
\label{cut:jj-iso}
\ee
The requirements summarized  in  Eqs. (\ref{cut:pT}--\ref{cut:jj-iso}) constitute our {\em acceptance cuts}. In order to calculate the production cross-section, simulate subsequent decays and detector resolutions and impose acceptance cuts, we have used a parton-level Monte-Carlo computer code. Pair and associated production of $\Delta^{\pm\pm}$ and $\Delta^{\pm\pm\pm}$ are simulated for $m_\Delta=500$ GeV and characteristic distributions are presented in the following. For simplicity, we have considered same mass for all the components of the quadruplet. 

\begin{figure}[htb]
\begin{center}
\includegraphics[width=12cm, height=8cm]{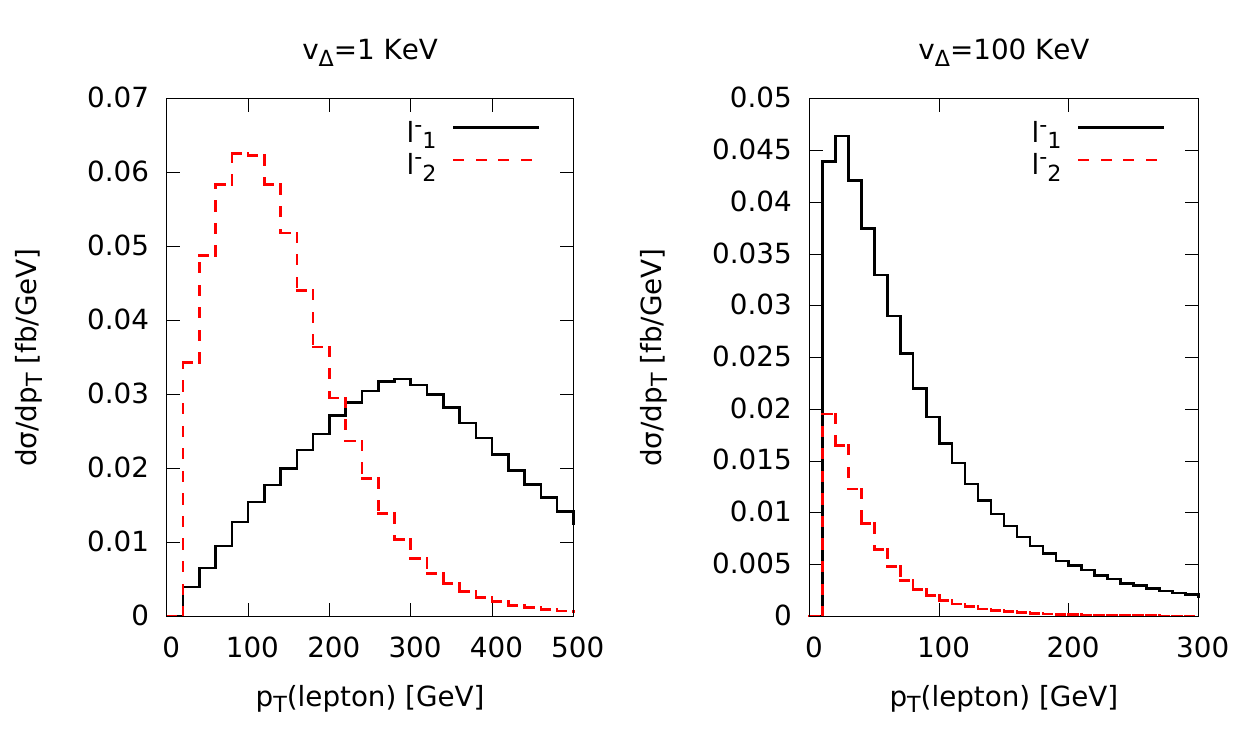}
\caption {Transverse momentum ($p_T$) distributions of hardest and second-hardest same-sign lepton after ordering the leptons according to their $p_T$ hardness ($p_T^{l^-_1}>p_T^{l_2^-}$) for small (left panel) and large (right panel) $v_\Delta$. $m_\Delta=500$ GeV is assumed.} 
\label{PT_lep}
\end{center}
\end{figure}

In Fig.~\ref{nlep}, we have presented the lepton multiplicity distributions for small (left panel) and large (right panel) $v_\Delta$. Fig.~\ref{nlep} clearly shows that lepton multiplicity varies between 0 to 6 for both small and large $v_\Delta$. For small $v_\Delta$, dileptons and 4-leptons multiplicity final states dominates over the others and interestingly, most of the dileptons are of same-sign. It is important to note that for small $v_\Delta$, the dominant decay modes of $\Delta^{\pm\pm}$ and $\Delta^{\pm\pm\pm}$ are $l^\pm l^\pm$ and $l^\pm l^\pm W^\pm$, respectively. Therefore, pair and associated production of $\Delta^{\pm\pm}$ and $\Delta^{\pm\pm\pm}$ always result atleast 4-leptons (including taus) in final state. Five and six leptons arise when $W$-decays leptonically. Dileptons arise when a pair of taus from the decay of $\Delta^{\pm\pm}$ or $\Delta^{\pm\pm\pm}$ decays hadronically. Since, dileptons signature is a consequence of $\tau$-hadronic decay and the decay of $\Delta^{\pm\pm}$ and $\Delta^{\pm\pm\pm}$ into leptons are flavor conserving, majority of dileptons are same-sign dileptons for small $v_\Delta$. Small number of events with opposite sign dileptons (OSD) arise from the $\Delta^{\pm\pm\pm}\Delta^{\mp\mp\mp}$ production followed by $\Delta^{\pm\pm\pm}\to\tau^\pm \tau^\pm W^\pm$ as well as $\tau$-hadronic and $W$-leptonic decay. On the other hand, for large $v_\Delta$, $\Delta^{\pm\pm}$ and $\Delta^{\pm\pm\pm}$ dominantly decays to $W$-bosons and subsequent leptonic decays of $W$-bosons give rise to leptonic final states. Therefore, in this case higher lepton multiplicity states are suppressed by the leptonic branching ratios of $W$-boson as can be seen from Fig.~\ref{nlep} (right panel). Moreover, in this case all the dileptons are not necessarily same-sign dileptons as in the case of small $v_\Delta$. However, there is a significant amount of SSD and SS3L for both small and large $v_\Delta$. In Fig.~\ref{njet}, we have presented the parton level jets multiplicity distributions for small (top panel) and large (bottom panel) $v_\Delta$. As expected for small $v_\Delta$, jet multiplicities are usually small. Whereas, for  large $v_\Delta$, we have large jet multiplicity final states. However, it is important to note that our computation is done at parton level without incorporating initial state radiation/final state radiation (ISR/FSR). Inclusion of ISR/FSR jets would significantly change the shape of jet multiplicity distributions in Fig~\ref{njet}.

\begin{figure}[htb]
\begin{center}
\includegraphics[width=12cm, height=8cm]{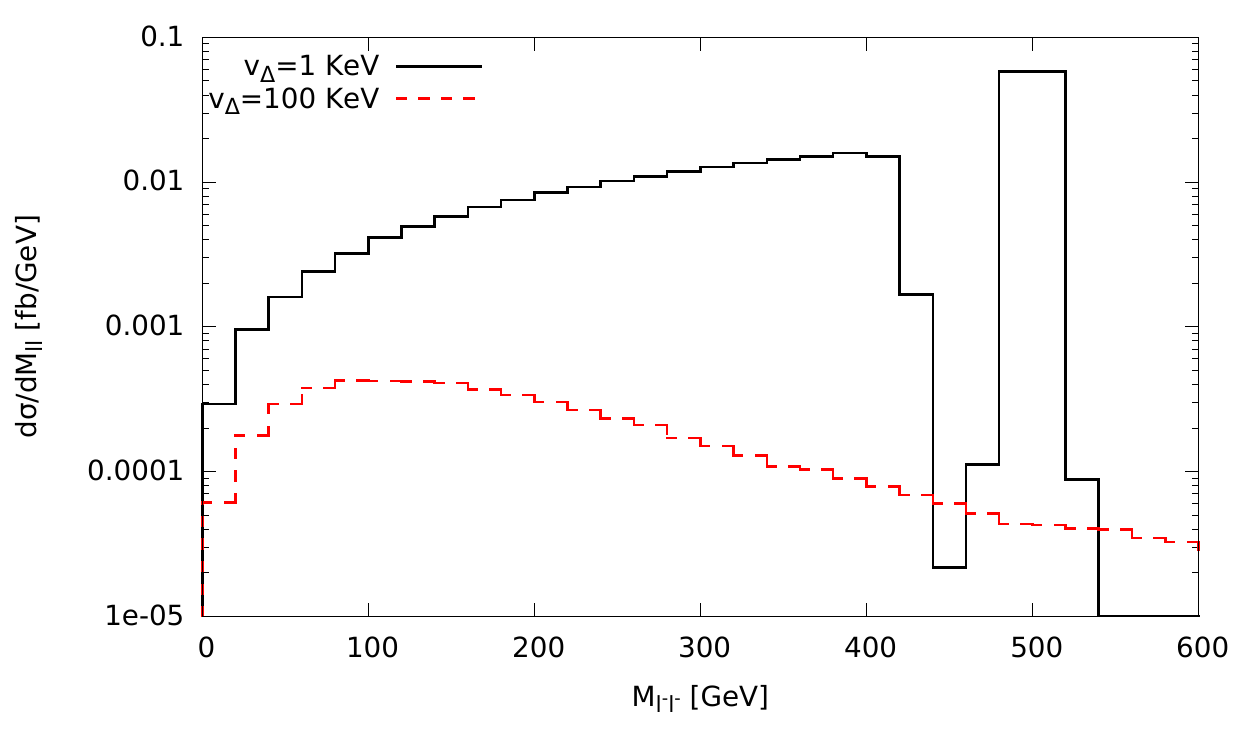}
\caption {Invariant mass distributions of same-sign lepton pairs after the acceptance cuts in Eqs. (\ref{cut:pT}--\ref{cut:jj-iso}). In the simulation of the same-sign lepton pairs invariant mass distribution, we have considered both the pair and associated production of triply and doubly scalars namely, $pp \to \Delta^{+++}\Delta^{---},~\Delta^{++}\Delta^{--},~\Delta^{+++}\Delta^{--}~{\rm and}~\Delta^{++}\Delta^{---}$}. 
\label{M_lep}
\end{center}
\end{figure}

In Fig.~\ref{PT_lep}, we have presented the transverse momentum ($p_T$) distributions of hardest and second-hardest same-sign leptons after ordering the leptons according to their $p_T$ hardness ($p_T^{l^-_1}>p_T^{l_2^-}$). Left and right panel in Fig.~\ref{PT_lep} corresponds to small and large $v_\Delta$, respectively. For small $v_\Delta$, 500 GeV $\Delta^{\pm\pm}(\Delta^{\pm\pm\pm})$ directly decays to a same-sign lepton pairs (leptons pairs plus $W$-boson) and hence, the lepton transverse momentum in this case are usually large. However, for large $v_\Delta$, leptons arise from the decay of the $W$-boson. As a result, the leptons are soft for large $v_\Delta$ as can be seen from Fig.~\ref{PT_lep} (right panel). Moreover, the possibility of getting a second lepton with same-sign is small for large $v_\Delta$ (see Fig.~\ref{nlep}).

For small $v_\Delta$, the doubly charge quadruplet scalar decay into a pair of same-sign leptons. Therefore, the characteristic signature for small $v_\Delta$ is a peak in the invariant mass distribution of same-sign leptons. We have considered events with 4-leptons with two positively and two negatively charged leptons and plotted the invariant mass distribution of same-sign dilepton pairs in Fig.\ref{M_lep}. A invariant mass peak at $500$ GeV is clearly visible in Fig.\ref{M_lep}. It is interesting to notice that the characteristic $\Delta^{\pm\pm}$ invariant mass peak is accompanied by a nearby invariant mass edge. The SSD invariant mass edge at ($m_\Delta-m_W$) for small $v_\Delta$ results from the decay of $\Delta^{\pm\pm\pm}$ into same-sign lepton pairs and a $W$-boson. Therefore, for small $v_\Delta$, the characteristic signature of quadruplet scalars in the framework of this model is a SSD invariant mass peak (at $m_\Delta$) accompanied by a nearby invariant mass edge (at $m_\Delta-m_W$). The search for the invariant mass peak in the same-sign dilepton  invariant mass distribution is the most promising channel for the discovery of small $v_\Delta$ region of the parameter space. The ATLAS and CMS collaborations of the LHC experiment are actively studying  same-sign dilepton invariant mass distributions. In absence of any significant deviation from the SM background prediction at the ATLAS detector, we have already extracted a bound of about 725 GeV on $M_{\Delta^{\pm\pm}}$ in the previous section. With more data, the LHC will be able to probe larger $M_{\Delta^{\pm\pm}}$ regions and observation a invariant mass edge in association with the characteristic SSD invariant mass peak will surely indicate towards a underlying physics model of present kind. However, for large $v_\Delta$, the invariant mass distribution of same-sign lepton pairs do not show any characteristic feature. Moreover, as can be seen from Fig.~\ref{nlep} and Fig.~\ref{PT_lep}, large $v_\Delta$ is corresponding to  suppressed and softer multi-leptons in the final state and hence, making the collider phenomenology challenging. The signatures and LHC discovery reach of large $v_\Delta$ part of parameter space is discussed in the following.

\subsubsection{The LHC discovery reach for large $v_\Delta$}
 
The high lepton multiplicity final states namely, 4-leptons, 5-leptons and 6-leptons states, are suppressed by $W$-boson leptonic branching ratios for large $v_\Delta$. However, there are significant amount of dileptons and 3-leptons events. Dileptons and 3-leptons final states suffer from huge SM backgrounds from top-antitop, $\gamma/Z/W$-boson production. However, it is important to note that $t \bar t$ and $\gamma/Z/W$-boson productions dominantly give rise to leptons with opposite charges and the SM contributions to SSD and SS3L are very small or negligible. On the other hand, the signal SSD and SS3L are suppressed (see Fig.~\ref{nlep} right panel) compared to total 2L and 3L final states only by some factor (in particular, by a factor of 2.5 and 10 for SSD and SS3L, respectively). In view of this facts, we have considered SSD and SS3L for further study.

\begin{figure}[htb]
\begin{center}
\includegraphics[width=7.5cm, height=8cm]{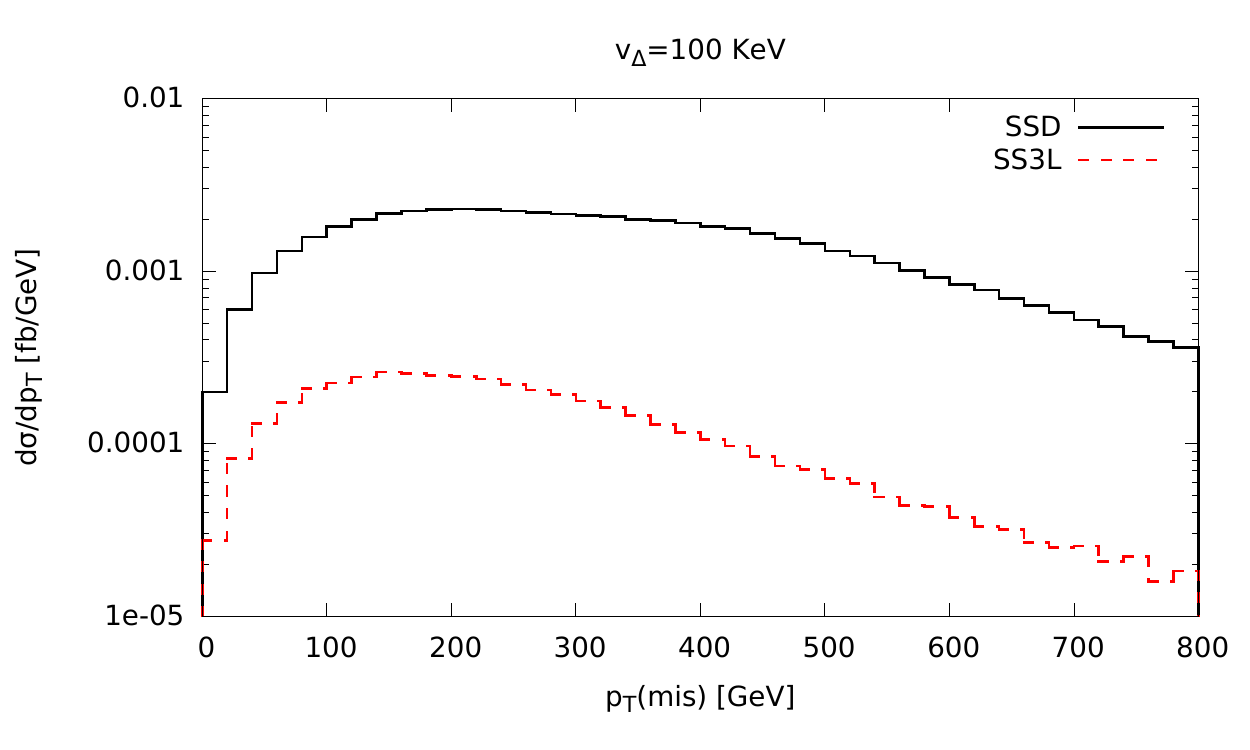}
\includegraphics[width=7.5cm, height=8cm]{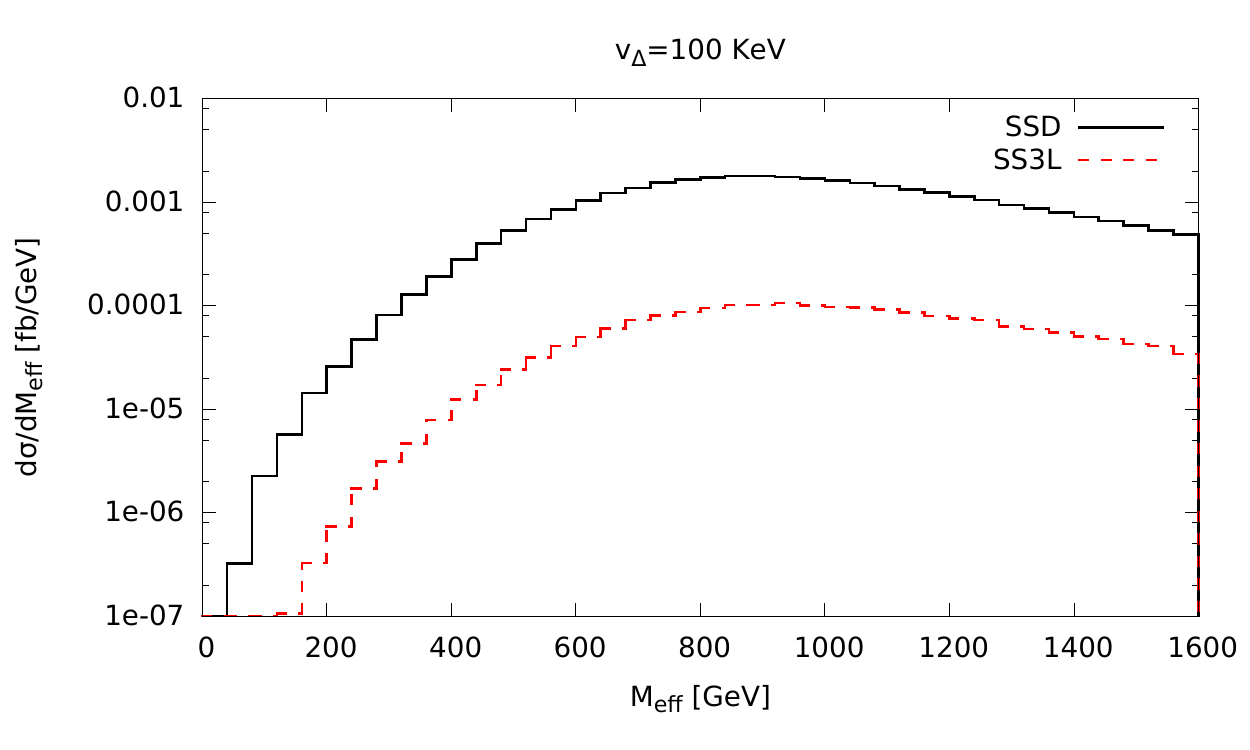}
\caption {Missing transverse momentum, $p_T\!\!\!\!\!\/~$, (left panel)  and effective mass, $M_{eff}$ (right panel) distributions for SSD and SS3L events after the acceptance cuts.} 
\label{dis_cut}
\end{center}
\end{figure}

We have selected events with exactly 2- and 3-leptons with same electric charge for further analysis. For large $v_\Delta$, the lepton arises from the $W^\pm\to l \nu$ decay. The resulting neutrino remains invisible in the detector and gives rise to missing transverse momentum ($p_T\!\!\!\!\!\/~$) signature. The missing transverse momentum defined in terms of the 
total visible momentum, as,
\be
\not p_T \equiv \sqrt{ \bigg(\sum_{\rm vis.} p_x \bigg)^2 
                 + \bigg(\sum_{\rm vis.} p_y \bigg)^2 }.\nonumber
\ee  
Therefore, the leptonic final states for large $v_\Delta$ are always accompanied by some amount of missing transverse momentum. In Fig.~\ref{dis_cut}, we have presented the missing transverse momentum distributions for SSD and SS3L events after the acceptance cuts. Fig.~\ref{dis_cut} (right panel) corresponds to the effective mass ($M_{eff}$) distributions where $M_{eff}$ is defined as the scalar sum of the $p_T$ of the signal leptons and jets as well as $p_T\!\!\!\!\!\!/~$.

\begin{figure}[htb]
\begin{center}
\includegraphics[width=12cm, height=8cm]{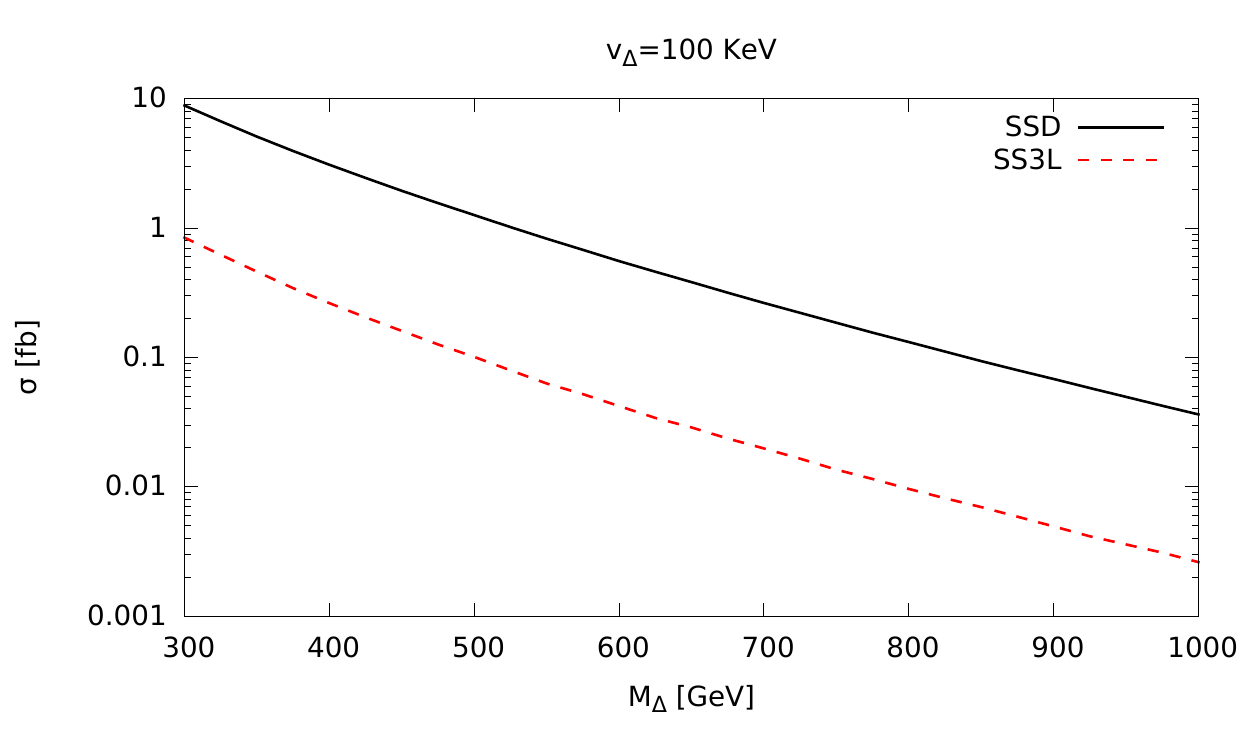}
\caption {Signal SSD and SS3L cross-sections after the selection cuts as a function of quadruplet mass.} 
\label{cross_cut}
\end{center}
\end{figure}

In the SM, same sign dilepton and tri-lepton arise mainly from the production of $t\bar t W^\pm$ and multiple gauge bosons ($W$ and/or $Z$) productions. $t\bar t W^\pm$ contributes to SSD when $t(\bar t)$ decays leptonically, $\bar t(t)$ decays hadronically and $W^{+(-)}$ decays leptonically. On the other hand, $ZW^\pm$ contributes to SSD when both $Z$ and $W$ decays leptonically and one lepton from $Z$-decay falls out side the coverage of the detector ($p_T<20$ GeV and/or $|\eta|>$2.5) or do not identified as individual entities ($\Delta R_{ll}<0.4$ or $\Delta R_{lj}<0.4$). These backgrounds ($t\bar t W^\pm$ and dibosons) fall in the category of irreducible backgrounds since these SM processes contains two same-sign prompt leptons or at least three prompt leptons out of which one lepton falls out side detector coverage. Contribution to SSD may also arise from events containing electrons with mismeasured charges, mainly from the production of top quark pairs, and events containing at least one fake or non-prompt lepton. The fake or non-prompt lepton mainly originates from heavy-flavour hadron decays in events containing top quarks, or $W$ or $Z$ bosons. For example, production of $t\bar t$ pairs may contribute to SSD when  $t\bar t$ pairs decays semileptonically and the $b$-quark from the hadronically decaying top decays into a lepton. These backgrounds fall into the category of reducible backgrounds because the lepton from the heavy-flavour hadron decays is always accompanied by  lots of hadronic activities around it or a jet within close proximity of the lepton and thus, stronger lepton isolation cuts can be used to reduce these backgrounds. The SM background contribution to SSD was studied by ATLAS collaboration \cite{ATLAS_SSD1} in the context of 13 TeV LHC. In order to reduce the SM background contribution to SSD + $\not p_T$, we have used ATLAS suggested cuts on $\not p_T>125$ GeV and $m_{eff}>650$ GeV as {\em selection cuts}. With these set of event selection criteria, dominant SM contribution to the SSD arises from $ZW$ and $t\bar tW$ production. We have simulated  $ZW$ and $t\bar tW$ in association with upto 3 and 4 additional jets, respectively, using ALPGEN \cite{Mangano:2002ea} and the resulting SSD background cross-section after the selection cuts is estimated to be 0.6 fb at the LHC with 13 TeV center of mass energy. 

On the other hand, there is no irreducible source of SS3L in the SM. The contribution to SS3L may arise from $t\bar t,~t\bar t W^\pm,~t\bar t b \bar b,~t \bar t t \bar t$ e.t.c.  production when one (only for $t \bar t W$) or few (all other sources) lepton(s) from heavy-flavour hadron decays are identified as isolated leptons. As discuss in the previous paragraph, lepton isolation cuts significantly reduce this background.  Dominant contribution to SS3L background arises from $t \bar t W$, when one top and $W$ decay leptonically and result into like sign leptons and the third like sign lepton comes from the leptonic decay of $b$ hadrons.  We have introduced the following selection cuts to study the SS3L signature.
\begin{itemize}
\item $p_T^{l_1}>30$ GeV, $p_T^{l_2}>30$ GeV, $p_T^{l_3}>20$ GeV and $\not p_T>50$ GeV.
\end{itemize}  
For SS3L background, one or more leptons arise from the decay of heavy-flavour hadrons which can not be simulated in the framework of parton level Monte-Carlo analysis. Therefore, we have used PYTHIA (v6.4.28) \cite{pythia} to simulate $t\bar t W$ production, subsequent decays and hadronization. To reconstruct the jets, we have used FastJet anti-$k_t$ algorithm \cite{antikt} implemented in Fastjet package \cite{fastjet} with a cone of $\Delta R=0.4$ and minimum transverse momentum of 20 GeV. Lepton isolation criteria plays a crucial role for SS3L background. For a isolated lepton, we demand $\sum p_T {\rm (hadron)} /p_T {\rm (lepton)}\le 0.2$, where the sum is over all hadrons within a cone of $\Delta R \le 0.2$ around the lepton. Other object reconstruction criteria listed in Eqs. (\ref{cut:pT}--\ref{cut:jj-iso}) are applied subsequently. With these set of event selection cuts, we have estimated the $t\bar t W$ contribution to the SS3L to be less than $10^{-3}$ fb. Therefore, there will be no SS3L background events with the above mentioned set of cuts upto 1000 fb$^{-1}$ integrated luminosity. The signal SSD and SS3L cross-sections after the selection cuts are presented in Fig.~\ref{cross_cut}.

\begin{figure}[htb]
\begin{center}
\includegraphics[width=12cm, height=8cm]{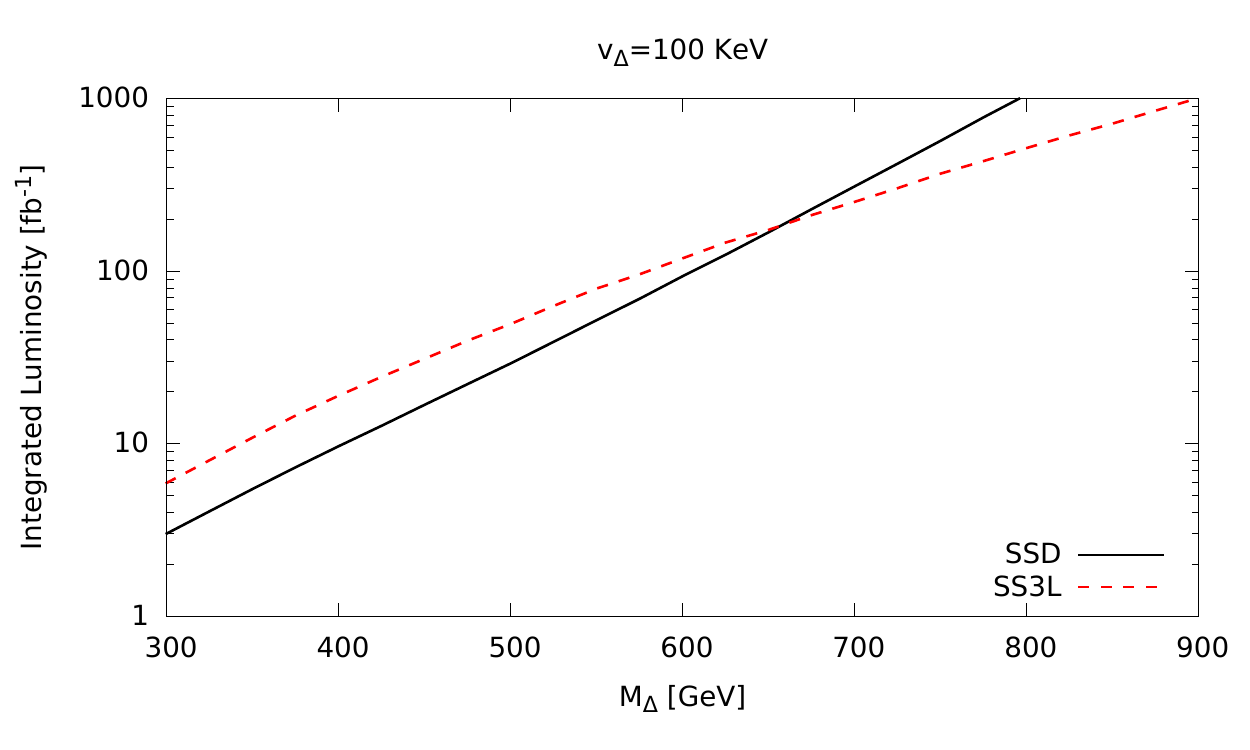}
\caption {Required luminosity at the 13 TeV LHC for 5$\sigma$ discovery of quadruplet scalars with large $v_\Delta$ as a function $M_\Delta$.} 
\label{reach}
\end{center}
\end{figure}

In order to calculate the discovery reach of the LHC with 13 TeV center of mass energy, we define the signal to be observable over a non-zero background for an integrated luminosity ${\cal L}$ if,
\begin{equation}
\frac{N_{S}}{\sqrt{N_B+N_S}} \ge 5,
\end{equation}
where, $N_{S(B)}=\sigma_{S(B)} {\cal L}$, is the number of signal (background) events for an integrated luminosity ${\cal L}$. However, if the number of background event is less than one for a integrated luminosity ${\cal L}$ (as in the case of SS3L), then we demand 5 signal event for the discovery. In Fig.~\ref{reach}, we have presented required luminosity of the 13 TeV LHC for 5$\sigma$ discovery of quadruplet scalars with large $v_\Delta$ as a function of quadruplet mass. Fig.~\ref{reach} shows that for lower $M_\Delta$, SSD is the promising channel however, for $M_{\Delta}>650$ GeV, SS3L becomes promising.

\subsubsection{Collider implications of vector like leptons}

 Beside charged scalars, another distinctive feature of this model is the prediction of doubly and singly charged leptons at the TeV scale. However, tiny neutrino masses, generated dominantly via tree level effective dimension-7 operators, require triplet fermions masses to be at the range of few TeVs (see Fig.~\ref{fig:contour} and Table~\ref{numassorder}). The electroweak pair production cross-sections TeV mass triplet fermions are miniscule at LHC with 13 TeV center of mass energy. It is needless to mention that either masses of the vector like leptons or masses of the quadruplet scalars have to be heavy (at TeV scale) to give correct order neutrino mass as shown in Eq. \ref{eq:mnu2}. Otherwise, we have to assume extremely small tiny Yukawa coupling to compensate that and we are not concentrating on that scenario. Here, in this work, we have mainly studied the production and signatures of the quadruplet scalars, in particular, multi-charged  quadruplet scalars at the LHC. For completeness of the study, here we discuss collider implication of the triplet vector-like fermions in the complimentary regions.

\begin{figure}[htb!]
\begin{center}
\includegraphics[width=13cm, height=8cm]{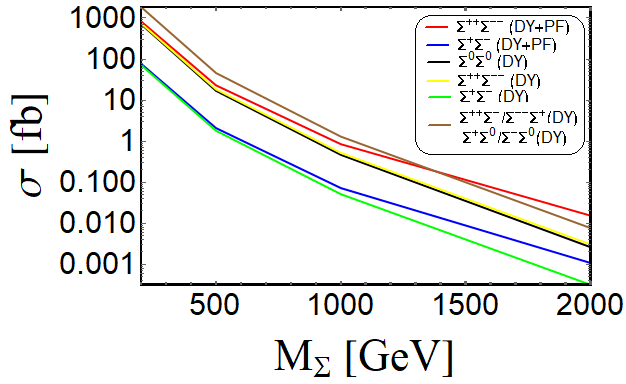}
\caption {Pair production cross-section of vector like leptons $(\Sigma^{++},\Sigma^{+},\Sigma^{0})$ at 13 TeV LHC.} 
\label{crosslep}
\end{center}
\end{figure}

\begin{table}[htb!]
\Huge
	\centering
	\scalebox{0.46}{
		\begin{tabular}{|c||c|c|c|c|c|c|} \hline
	              & $\Sigma^{++} \to \lp \wbp$ & $\Sigma^{+} \to \nub \wbp$ & $\Sigma^{+}  \to \lp Z$ & 
	                $\Sigma^{+} \to \lp H$ & $\overline{\Sigma^0} \to \nub Z$ & $\overline{\Sigma^0} \to \nub H$  
\\\hline
\hline
			$\Sigma^{--} \to \lm \wbm$&
			$ \lm \lp \wbm \wbp$&
			$ \lm \nub \wbm \wbp$&
			$ \lm \lp  \wbm Z$&
			$ \lm \lp \wbm  H $&
			-& 
			-

\\\hline
				$\Sigma^{-} \to \nu \wbm$&

			$\nu \lp \wbm \wbp $&
			$\nu \nub \wbm \wbp $&
			$\nu \lp \wbm Z $&
			$\nu \lp \wbm H $&
			$\nu \nub \wbm Z $&
			$\nu \nub \wbm H $
\\\hline
			$\Sigma^{-}  \to \lm Z$ &
			$\lm \lp Z \wbp$ &
			$\lm \nub Z \wbp$ &
			$\lm \lp Z Z$ &
			$\lm \lp Z H$ &
			$\lm \nub Z Z$ &
			$\lm \nub Z H$ 

\\\hline
			$\Sigma^{-} \to \lm H$ &
			$ \lm \lp H \wbp$ &
			$ \lm \nub H \wbp $ &
			$ \lm \lp H Z $ &
			$ \lm \lp H H$ &
			$ \lm \nub H Z $ &
			$ \lm \nub H H$ 

\\\hline
			$\Sigma^0 \to \nu Z$&
			-&
			$ \nu \nub Z \wbp$ &
			$ \nu \lp Z Z$ &
			$ \nu \lp Z H$ &
			$ \nu \nub Z Z$ &
			$ \nu \nub Z H$ 

\\\hline
			$\Sigma^0 \to \nu H$&
			-&
			$ \nu \nub H \wbp$ &
			$ \nu \lp H Z$ &
			$ \nu \lp H H$ &
			$ \nu \nub H Z$ &
			$ \nu \nub H H$ 
\\\hline
		\end{tabular}
		}
	\caption{\footnotesize Exotic lepton decay channels to SM particles along with the final state signatures of pair/associated production.} 
	\label{channels}
\end{table}

At the LHC, $\Sigma^{\pm\pm} \Sigma^{\mp\mp}$, $\Sigma^{\pm} \Sigma^{\mp}$ and $\Sigma^{0} \Sigma^{0}$ are pair produced via the s-channel $\gamma$ and/or Z exchanges. In addition to that, photon initiated processes also significantly contribute to the pair production of the singly and doubly charged leptons at the LHC. Pair and associated production cross-sections of vector like leptons $(\Sigma^{++},\Sigma^{+},\Sigma^{0})$ at 13 TeV LHC is shown in Fig. \ref{crosslep}. Due to the charge enhancement factor of 16, doubly charged lepton $\Sigma^{++}$  is largely pair produced compared to $\Sigma^{\pm}$ and $\Sigma^{0}$. DY pair production of $\Sigma^{0}\Sigma^{0}$ gets contribution only from $Z$-boson exchange in the $s$-channel. Whereas, both photon and $Z$-boson exchange in the $s$-channel contributes to DY production of $\Sigma^{+}\Sigma^{-}$ pairs. It is important to mention that coupling strength of doubly charged and neutral leptons with Z boson is large compared to the singly charged lepton. Moreover, there is a destructive interference between photon and $Z$-boson exchange Feynman diagrams for $\Sigma^{\pm\pm}$ pair production. Being triplet under $SU(2)_L$, $\Sigma^{++},~\Sigma^{+}~{\rm and}~\Sigma^{0}$ are degenerate at tree level. This degeneracy is removed by the radiative corrections. However, the mass splitting between $\Sigma^{++},~\Sigma^{+}~{\rm and}~\Sigma^{0}$ results from the radiative corrections are expected to be small.  

After being pair produced, the heavy triplet leptons decay into  the SM particles. The decay modes of heavy triplet leptons crucially depend on the hierarchy between $M_\Sigma$ and $M_\Delta$. For  $M_\Delta<M_\Sigma$, triplet leptons dominantly decay into quadruplet scalars in association with a SM lepton.
\begin{eqnarray}
 \Sigma^{++} \to \Delta^{+++}l^{-}, \Delta^{++}\nu;
 \\
 \Sigma^{+} \to \Delta^{++}l^{-}, \Delta^{+}\nu;
 \\
 \Sigma^{0} \to \Delta^{+}l^{-}, \Delta^{0}\nu.
\end{eqnarray}
The subsequent decay of quadruplet scalars and their collider signatures are already been discussed in details in the previous section. On the other hand, for $M_\Delta>M_\Sigma$, the triplet leptons can decay into a SM lepton (both charged or neutral) in association with a SM EW gauge boson ($W$ or $Z$-boson) or Higgs boson. The decay into $W$ or $Z$-boson arises due to the Yukawa interactions in Eq.~\ref{Lsig} which induce small mixing between the lepton triplets and usual SM doublets. These decay modes are listed in Table \ref{channels} along with the final state signatures of pair/associated production of the exotic leptons. Due to the small splitting between the masses of triplet leptons, the heavier triplet leptons can decay into the lighter one in association with a off-shell $W$-boson which subsequently decays leptonically or hedronically giving rise to very soft leptons or hadrons at the LHC. However, it is important to note that these decays are tree level 3-body decays and hence, suppressed by the $W$-boson mass. A detailed collider study of the triplet vector like fermions is beyond of the scope of this study. However, the final states (listed in Table \ref{channels}) resulting from the pair/associated production of exotic leptons give rise to interesting multi-leptons signatures which require a detailed study at the high energy (HE) and/or high luminosity (HL) LHC.

\color{black}


\section{Summary and Discussions} \label{sec:4}
In this article, we have presented a model, which can generate small neutrino masses via dimension seven effective operators $L L HH(H^\dagger H)/M^3$ and can also be probed at the LHC through the multi-lepton signatures. We have investigated the visibility of the triply and doubly charged scalars at the LHC. We have found that the photon photon fusion also contributes to pair production process at a significant level at the LHC due to the substantially enhanced electromagnetic coupling. This, we emphasize in this literature, is comparable to the DY channel, and must be included in a complete and accurate estimate. We consider the spectacular  multi-lepton final states driven by the decay of the $\Delta^{\pm\pm\pm}(\Delta^{\pm\pm})$ into same sign trileptons (dileptons). These channels not only lead to remarkably background-free signatures of the doubly charged scalars, but also can demonstrate a crucial link between observations at high energy colliders and the discussed mechanism of neutrino mass generation.

The characteristic collider signatures of quadruplet scalars crucially depend on the decayes of these scalars and hence, on the value of the induced VEV, $v_\Delta$. For small $v_\Delta$, production and decay of quadruplet scalars gives rise to a same-sign dilepton invariant mass peak at $m_{\Delta^{\pm\pm}}$ which is accompanied by a invariant mass edge at $m_{\Delta^{\pm\pm\pm}}-m_W$. In absence of any significant deviation in the LHC same-sign dilepton invariant mass data, we derived a bound of about 725 GeV on  $m_{\Delta^{\pm\pm}}$. On the other hand, for large $v_\Delta$, the pair and associated production of $\Delta^{\pm\pm}$ and/or $\Delta^{\pm\pm\pm}$ give rise to softer leptons in the final states with suppressed cross-sections. We have studied SSD and SS3L final states as signatures of quadruplet scalars for large $v_\Delta$. We found that the LHC with 13 TeV center of mass energy and 100 inverse femtobarn integrated luminosity will be able to probe $M_\Delta$ upto 600 GeV. We also briefly discussed the signature of TeV scale triplet fermions at the LHC. A detailed collider phenomenology of the triplet fermions seems interesting at the HE-LHC and/or HL-LHC.

\section*{Acknowledgement}
The work of SN and SJ was in part supported by US Department of Energy Grant Number DE-SC 0016013. KG acknowledges support from the Department of Atomic Energy, Government of India, via Inspire Faculty Project. The work of SJ was also supported in part by the Fermilab Distinguished Scholars Program.

\end{document}